\documentclass[12pt]{article}
\usepackage{amssymb,amsmath}
\usepackage{graphicx}
\usepackage[arrow,matrix,curve]{xy}

\usepackage{amscd}
\usepackage{amsfonts}
\usepackage{amsthm}
\usepackage{mathrsfs}

\newcommand{\e}{{\rm e}}
\newcommand{\ii}{{\rm i}}
\newcommand{\eqn}[1]{(\ref{#1})}

\newcommand{\bea}{\begin{eqnarray}}
\newcommand{\eea}{\end{eqnarray}}
\def\beqa{\begin{eqnarray}}
\def\eeqa{\end{eqnarray}}
\def\nn{\nonumber}

\newcommand{\eq}{\begin{equation}}
\newcommand{\eqa}{\begin{eqnarray}}
\newcommand{\en}{\end{equation}}
\newcommand{\ena}{\end{eqnarray}}

\def\sk{\vskip .4cm}
\def\noi{\noindent}
\def\nn{\nonumber}
\def\epsi{{\varepsilon}}
\def\f{{\rm{f}\,}}

\newcommand{\om}{\omega}
\newcommand{\Om}{\Omega}
\newcommand{\al}{\alpha}
\newcommand{\la}{\lambda}
\newcommand{\ga}{\gamma}
\newcommand{\Ga}{\Gamma}
\newcommand{\de}{\delta}

\newcommand{\RR}{{\mathcal R}}
\newcommand{\oR}{{\bar{\R}}}
\newcommand{\st}{{\star}}

\newcommand{\UU}{{U}\Xi}
\newcommand{\FF}{\mathcal F}
\newcommand{\varepsi}{\varepsilon}

\newcommand{\AAs}{{A_\st }}

\newcommand{\Xis}{{\Xi_\st }}
\newcommand{\Oms}{\Omega_\st}

\newcommand{\D}{\Delta}
\newcommand{\TT}{{\mathcal T}}
\newcommand{\lls}{\ll^\st }

\newcommand{\re}{{\rangle}}

\newcommand{\g}{\mathsf{g}}
\newcommand{\ots}{\otimes_\st}
\newcommand{\res}{\re_\st}

\newcommand{\LL}{\mathcal{L}}

\newcommand{\x}{\mathsf{x}}
\newcommand{\y}{\mathsf{y}}
\newcommand{\cc}{\mathbb{C}}

\newcommand{\ff}{{\sf f}}
\newcommand{\rr}{{\R}}

\def\ll{{\mathcal L}}
\def\d{{\rm d}}
\def\D{\Delta}
\def\st {\star}
\def\dd{{\triangledown}}
\def\dds{\triangledown^\st}

\def\be{\beta}

\def\of{{\bar{{\rm{f}}\,}}}
\def\R{{R}}
\def\oR{{\bar{\R}}}
\def\cc{\mathbb{C}}

\def\TT{{\mathcal T}} 
\def\Xis{{\Xi_\st }}
\def\ots{\otimes_\st}
\def\rr{\mathsf{R}^\st}
\def\tr{{\mathsf{T}}^\st}
\def\ric{{\mathsf{Ric}}^{\st}}
\def\rrc{\mathsf{R}}
\def\trc{{\mathsf{T}}}
\def\ricc{{\mathsf{Ric}}}
\def\G{{\mathsf{G}}}

\def\Oms{\Omega_\st}
\def\Om{\Omega}
\def\le{{\langle}}

\begin{document}
\begin{titlepage}
\begin{flushright}

\baselineskip=12pt
DISTA/FIS - 018/09\\
\hfill{ }\\
\end{flushright}

\begin{center}

\baselineskip=24pt

{\Large\bf Noncommutative Gravity Solutions}

\baselineskip=14pt

\vspace{1cm}

{\bf Paolo Aschieri$^{1,2}$, Leonardo Castellani$^2$}
\\[6mm]
{\it $^{1}$Centro Studi e Ricerche ``Enrico Fermi'' Compendio Viminale, 00184 Roma, Italy}\\[.4em]
  {\it $^{2}$Dipartimento di Scienze e Tecnologie
 Avanzate, Universit\`{a} del
 Piemonte Orientale,}\\ and {\it INFN, Sezione di Torino, gruppo collegato di Alessandria }\\[.4em]
{\small{{aschieri@to.infn.it,  leonardo.castellani@mfn.unipmn.it}}}
\\[10mm]

\end{center}

\vskip 2 cm

\begin{abstract}
We consider noncommutative geometries obtained from a triangular
Drinfeld twist and review the formulation of  noncommutative gravity.
A detailed study of the abelian twist geometry is presented, including the fundamental 
theorem of noncommutative Riemannian geometry.
Inspired by \cite{Schupp,SchenkelOhl}, we obtain  solutions of noncommutative 
Einstein equations by considering twists that are compatible with the 
curved spacetime metric. 
\end{abstract}

\end{titlepage}

\tableofcontents
\section{Introduction}
The study of solutions of noncommutative field theories is of primary importance for   a better understanding of the theories themselves and in discussing their phenomenological implications. 
Some of the relevant literature on noncommutative gravity solutions 
can be found in references 
\cite{stargravitysol1}, 
\cite{stargravitysol2}.
Of particular interest are noncommutative black hole solutions and
noncommutative cosmological solutions.
The study of solutions may also help to understand the relations between 
different approaches to noncommutative gauge and gravity theories. 
In the case of noncommutative gravity we mention
for example the metric approach of ref.s \cite{GR1,GR2} and the vielbein approach 
that allows coupling to fermions of ref.s \cite{AC1,AC2}.

We here study solutions of the  noncommutative Einstein equations considered in \cite{GR1, GR2}. The theory in \cite{GR1, GR2}  is obtained by demanding covariance 
under noncommutative general coordinate transformations and it is based on noncommutative Riemannian geometry. It can address local and also global  (topological) aspects in noncommutative general relativity and noncommutative geometry.
The theory therefore allows also for a detailed study of its solutions 
that goes beyond a qualitative analysis. This is object of the present paper.

Noncommutative Einstein equations are in principle quite hard to solve.
By expanding these equations in the noncommutativity parameter $\theta$
we see that they contain partial derivatives of any order, the order increasing with 
the power of $\theta$.
 One can consider approximate gravity solutions by truncating this expansion at a given order in 
$\theta$. This is physically sensible 
if we consider $\theta$ as a small perturbation to commutative spacetime.
On the other hand it is also important to present exact nonperturbative solutions. 
Here one approach, that has been fruitfully studied in noncommutative gauge theories, is to consider topological solutions, like for example gravitational instantons solutions.   
Our approach, inspired by \cite{Schupp, SchenkelOhl}, is to consider noncommutative gravity solutions with symmetry properties. 

In the  framework of \cite{GR1, GR2}, that formulates noncommutative 
gravity based on Drinfeld twists \cite{drinfeld}, we prove that, provided the 
twist is in part constructed with Killing vector fields,  undeformed gravity solutions 
are also noncommutative gravity solutions. 
In this paper we study 
the case of Einstein equations in vacuum with and without cosmological constant, i.e., we study  Einstein spaces.  Our results holds for metrics with arbitrary signature, in particular Euclidean and Lorentian.
We stress that the commutative and noncommutative Einstein equations are different and therefore the corresponding set of solutions are different. 
It is only by requiring compatibility between the twist and the metric that we are able to find a 
subset of solutions that is common to both the commutative and the noncommutative theories. 
Note that even in this case  noncommutative and commutative connections in general differ, and hence geodesic motions also differ.  It is interesting to investigate the physical implications of these differences.

The class of star products and noncommutative manifolds we consider is a
rather large class.
The examples in Section 3 include  
quantization of symplectic and also of Poisson structures.
The algebra of functions of the
noncommutative  torus, of the noncommutative spheres 
\cite{Connes-Landi} and of further noncommutative manifolds 
(so-called isospectral deformations) considered in 
\cite{Connes-Landi}, and in \cite{Dubois-Violette}, 
\cite{Gayral:2005ad}, is associated to a 
$\st$-product structure obtained via a triangular Drinfeld twist
(see
\cite{Sitarz} and, for the four-sphere in \cite{Connes-Landi},
see \cite{Varilly}, \cite{Aschieri-Bonechi}).   In these cases the twist is abelian and 
entirely constructed with Killing vector fields, we consider the more general case where the triangular Drinfeld twist only in part contains Killing vector fields (cf. eq. \eqn{legs}, \eqn{legs2}).
The star products we study are however not the most general ones, in 
particular they are a subclass of those associated with a 
quasitriangular structure \cite{drinfeld2}: on that noncommutative 
algebra of functions there is an action of the braid group, 
while in the triangular case  there is an action of the permutation group.

It is remarkable how far in the program of formulating a noncommutative
differential geometry one can go using triangular Drinfeld twists. 
The study of this
class of $\st$-products geometries are first examples that can uncover 
some common features of a wider class of noncommutative geometries.

\sk
We consider three kinds of twists, in order of increasing generality: 

\noi I. ~Moyal-Weyl twist (associated with the Moyal-Weyl $\st$-product),
$$\FF=e^{-{i\over 2}\theta^{\mu\nu}{\partial\over \partial x^\mu}
\otimes{\partial\over \partial x^\nu}}$$
\noi II. ~Abelian twist (socalled because the vectorfields $\{X_a\}$ 
are mutually commuting),
$$\FF=e^{-{i\over 2}\theta^{ab}{X_a}
\otimes{X_b}}$$
\noi III. ~General (triangular) Drinfeld twist $\FF$.
\sk
In Section 2 we recall the main results of metric noncommutative gravity
with Moyal-Weyl twist. Particular solutions are discussed in Section 3.
They are  found by listing all the  metrics $\g$ compatible with the twist $\FF$,
in the sense that the twist $\FF$  is constructed in part with
Killing vector fields of $\g$. Among these metrics those that are classically
Einstein metrics are also shown to be noncommutative Einstein metrics.

In Section 4 we study the Riemannian geometry of abelian twists on a smooth manifold $M$. We locally reduce this study  to the Riemannian geometry of the Moyal-Weyl twist and then properly glue the local results in order to obtain the global ones on the whole noncommutative manifold $M$.
Here too if the twist $\FF$ contains in part  Killing vector fields of a commutative Einstein metric $\g$ then $\g$ is also a noncommutative Einstein metric. The corresponding noncommutative Levi-Civita connection is also given, and it is different from the commutative one.

The differential and Riemannian geometry of a general triangular Drinfel twist is recalled in Section 5.
New results include a detailed study of the contraction operator.
Uniqueness of the Levi-Civita connection of a noncommutative (pseudo-)Riemannian manifold is also proven.
Finally in Section 6 twisted gravity solutions  are considered for general Drinfeld twists
(case III above) that are constructed with affine Killing vectors. 
Here, differently from Section 3 and Section 4 (case I and II above), on  one hand we require a stronger compatibility condition between the twist and the metric, 
the twist being fully (and not in part) constructed with affine Killing vectors;
on the other hand we relax the Killing condition by considering the wider class of affine or homotetic Killing vectors.  
 
\section{Gravity with Moyal-Weyl $\st$-noncommutativity}
\noi {\it Functions}. 
The star product that implements the 
$x^\mu\st x^\nu-x^\nu\st x^\mu=i\theta^{\mu\nu}$ 
noncommutativity is given by  
\eq
(h\st g)(x)=e^{{i\over 2}\theta^{\mu\nu}{\partial\over \partial x^\mu}
{\partial\over \partial y^\nu}}h(x)g(y)|_{x=y}
\en
where $h$ and $g$ are arbitrary functions. 
This star product between functions can be obtained from the usual 
pointwise product $(hg)(x)=h(x)g(x)$ via the action of a twist 
operator $\FF$ 
\eq\label{starprodf}
h\st g:=\mu\circ \FF^{-1}(h\otimes g)~,
\en
where $\mu$ is the usual pointwise product between functions, 
$\mu(f\otimes g)=fg$, and the twist operator and its inverse are
\eq\label{MWTW}
\FF=e^{-{i\over 2}\theta^{\mu\nu}{\partial\over \partial x^\mu}
\otimes{\partial\over \partial x^\nu}}~,~~~
\FF^{-1}=e^{{i\over 2}\theta^{\mu\nu}{\partial\over \partial x^\mu}
\otimes{\partial\over \partial x^\nu}}~.
\en

We shall frequently use the notation (sum over $\al$ understood)
\eq
\label{Fff0}
\FF=\f^\al\otimes\f_\al~~~,~~~~\FF^{-1}=\of^\al\otimes\of_\al~,
\en
so that
\eq\label{fhfg}
f\st g:=\of^\al(f)\of_\al(g)~.
\en

In \cite{GR1} we developed a noncommutative geometry based on the twist \eqn{MWTW}
and the associated Moyal-Weyl $\st$-product (see also \cite{book, AschieriCorfu}).
The twist allows to define $\star$-products betweeen functions and vector fields and more in general between tensor fields. We thus obtained a deformed differential and Riemannian geometry that we briefly summarize.

\noi {\it Vector fields}.
Partial derivatives act on vector fields $v=v^\nu\partial_\nu$ via the 
Lie derivative action
\eq\label{onlyconst}
\LL_{\partial_\mu}(v)=[\partial_\mu,v]=\partial_\mu(v^\nu)\partial_\nu~.
\en
Similarly to (\ref{fhfg}) the product 
$\mu : A\otimes \Xi\rightarrow \Xi$
between the space $A$ of functions and the space $\Xi$ of vector fields is deformed into the product 
\eq\label{fhfg2}
h\st v=\of^\al(h) \of_\al(v)~,
\en
where $\of_\al$ acts on vectors via the Lie derivatives 
$\LL_{\of_\al}$ (Lie derivative along products of elements of the Lie algebra of vector fields $\Xi$ are defined simply by $\LL_{uv\cdots}=\LL_u\LL_v\cdots$).
Since $\FF^{-1}=e^{{i\over 2}\theta^{\mu\nu}\partial_\mu
\otimes \partial_\nu}$,
we obtain 
\eq
h\st \partial_\rho=h\partial_\rho
\en
This is so because the twist $\FF$ acts trivially on the vector field $\partial_\nu$ (cf. \eqn{onlyconst}).
More in general we have
$h\st v=\of^\al(h) \of_\al(v)
=\of^\al(h)\of_\al(v^\rho) \partial_\rho=
(h\st v^\rho)\partial_\rho ~.
$

We denote the space of vector fields with this 
$\st$-multiplication  by $\Xi_\st$. 
As vector spaces $\Xi=\Xi_\st$, but $\Xi$ is an $A$-module while $\Xi_\st$ is 
an $A_\st$-module.

\noi {\it Tensor fields}.  Tensor fields form an algebra with 
the tensor product $\otimes$ (over the algebra of functions). We define $\TT_\st$ to be the noncommutative 
algebra of tensor fields. As vector spaces $\TT=\TT_\st$; the noncommutative and associative
tensor product is obtained as in (\ref{fhfg}) and (\ref{fhfg2}):
\eq\label{fhfg3}
\tau\otimes_\st\tau':=\of^\al(\tau)\otimes \of_\al(\tau')~.
\en
here too, as in \eqn{fhfg2}, $\of^\al$ and $\of_\al$ act on tensors via the Lie derivatives $\LL_{\of^\al}$ and
$\LL_{\of_\al}$.
Notice that the action of the twist on the one forms $dx^\rho$ is trivial
\eq
\tau\otimes_\st dx^\rho=\tau\otimes dx^\rho~,~~
dx^\rho\otimes_\st \tau'=dx^\rho\otimes\tau'~,
\en
in particular $h\st dx^\rho=dx^\rho\st h=hdx^\rho$. 
This is so because the Lie derivative  along the vectors $\partial_\nu$
entering the twist $\FF$ vanishes on $dx^\rho$, $\LL_{\partial_\nu} dx^\rho=0$.
\sk

Vector fields act on tensor fields via the Lie derivative, and they form the Lie algebra of infinitesimal (local) diffeomorphisms. The space of $\st$-vector fields $\Xi_\st$ has 
similarly a $\st$-Lie derivative action on tensor fields. We have a deformed Leibniz rule so that if  $\tau$ and $\tau'$ transform as tensors also $\tau\otimes_\st \tau'$ transforms as a tensor (and  the matrix $\theta^{\mu\nu}$ remains invariant). Correspondingly we have a $\st$-Lie algebra of vector fields or of deformed infinitesimal (local)  diffeomorphisms.

A covariant derivative $\dds_\mu$ that has the same Leibnitz rule as the $\st$-Lie derivative can then be naturally  constructed.
\sk
The $\st$-differential geometry formulae simplify if we use the basis
$\partial_\mu$ and the dual basis $dx^\mu$.
Following \cite{GR1} (see also \cite{book}) for
any (undeformed)  metric tensor
\eq
\g=\g_{\mu\nu} dx^\mu\otimes dx^\nu= dx^\mu\otimes_\st dx^\nu\st
\g_{\mu\nu}
\en
there exists  a unique metric compatible and torsionfree covariant derivative $\dds_u$.
The Christoffel symbols are defined by
\eq
\dds_\mu\partial_\rho=\Gamma^\st_{\mu\rho}{}^\nu\st\partial_\nu~,~~
\en
or equivalently by
$\dds_\mu dx^\nu=-dx^\rho\st\Gamma^\st_{\mu\rho}{}^\nu~.$
A derivation similar to the one of the undeformed case leads to the explicit expression
\eq\label{Chr}
\Ga^\st_{\mu\nu}{}^\rho={1\over 2}(\partial_\mu \g_{\nu\sigma}+
\partial_\nu \g_{\sigma\mu}-\partial_\sigma \g_{\mu\nu})\st \g^{\st \sigma\rho}~,
\en
where $\g^{\st \sigma\rho}$ is the $\st$-inverse metric
\eq
\g^{\st \sigma\rho}\st \g_{\rho\nu}=\delta_\nu^\sigma~,~~
\g_{\nu\rho}\st \g^{\st \rho\sigma}=\delta_\nu^\sigma~.
\en
The torsion tensor is 
$
\tr_{\mu\nu}{}^\rho=\Ga^\st_{\mu\nu}{}^\rho-\Ga^\st_{\nu\mu}{}^\rho$ and vanishes.
The curvature tensor is given by\footnote{The relation between the coefficients $\rrc_{\mu\nu\rho}{}^\sigma$ defined in  \cite{book, AschieriCorfu} 
and those in \eqn{curvcoeff}  is $\rrc_{\mu\nu\rho}{}^\sigma={\rr}^\sigma_{~\rho\mu\nu}~.$}
\eq
{\rr}^\sigma_{~\rho\mu\nu}=\partial_\mu\Ga^\st_{\nu\rho}{}^\sigma-
\partial_\nu\Ga^\st_{\mu\rho}{}^\sigma
+\Ga^\st_{\nu\rho}{}^\beta\st\Ga^\st_{\mu\beta}{}^\sigma-
\Ga^\st_{\mu\rho}{}^\beta\st\Ga^\st_{\nu\beta}{}^\sigma~.
\label{curvcoeff}
\en

As in the
commutative case the Ricci tensor\index{Ricci tensor} is a contraction of the curvature
tensor,
\eq
\ric_{\mu\nu}={\rrc^{\st\rho}_{~\mu\rho\nu}}\label{Riccicoeff}~.
\en
Finally the noncommutative version of Einstein equation
in vacuum and in the presence of a cosmological constant $c\in \mathbb{R}$ is
\eq\label{Einsteincoeff}
\ric_{\mu\nu}=c_{\:\!} \g_{\mu\nu}~.
\en
In \cite{GR2} we defined  connection, curvature and Ricci curvature 
in a more intrinsic geometric language (without using coordinates) 
and in the case of an arbitrary manifold with noncommutativity given 
by a Drinfeld twist. 
As we review in Section 5.5 the Einstein equation in vacuum and in the presence of a cosmological constant reads 
\eq\label{EinsteinGen}
\ric=c_{\:\!} \g~.
\en
Metrics $\g$ that satisfy \eqn{EinsteinGen} are called $\star$-Einstein metrics,
and $(M,\FF,\g)$ is a $\star$-Einstein space.
Of course, if there exist local coordinates where the twist assumes the canonical Moyal-Weyl form \eqn{MWTW}, then, in the open domain of the chart defined by these coordinates, the connection, curvature and Ricci curvature are equivalently given by expressions \eqn{Chr}, \eqn{curvcoeff}, 
\eqn{Riccicoeff}; and  \eqn{EinsteinGen} is equivalent to \eqn{Einsteincoeff}.

\section{Gravity solutions I: Moyal-Weyl $\st$-product}
In this section we consider the manifold $\mathbb{R}^4$ (or $\mathbb{R}^N$)
with canonical Moyal-Weyl $\star$-product, or equivalently we  work locally
in an open neighbourhood with  coordinates  $\{x^\mu\}$
that satisfy the Moyal-Weyl relations $x^\mu\st x^\nu-x^\nu\st x^\mu=i\theta^{\mu\nu}$
(the constant matrix $\theta^{\mu\nu}$ being possibly degenerate).

The class of noncommutative (NC) Einstein metrics we consider are obtained by 
selecting those Einstein metrics of commutative spacetime 
that are compatible with the twist.
\sk
\noi {\it Theorem 1.~} Given the twist
 ${\mathcal F}=\e^{-\frac{\ii}{2}\theta^{\mu\nu}
{\partial_\mu}\otimes{\partial_\nu}}$
and a  metric $\g$,

\noi {\it i}) If the  Killing Lie algebra $g_K$ of the metric $\g$ has the twist
compatibility property
\eq
\theta^{\mu\nu}
{\partial_\mu}\otimes\label{legs}
{\partial_\nu}\in \Xi\otimes g_K +g_K\otimes \Xi
\en
then the NC curvature and  the NC Ricci curvature of the NC Levi-Civita connection are the undeformed ones. 

\noi {\it ii)} If the metric $\g$ is  an Einstein metric 
then it is  also a NC Einstein metric.
\sk
\begin{proof}
{\it i)} We show that the star product disappears from the 
expressions for the NC curvature, Ricci tensor and Einstein equation. 
The  inverse metric $\g^{-1}$ 
is invariant under the same Lie algebra $g_K$ as $\g$
and hence coincides with the $\st$-inverse metric,
$\g^{\sigma\rho}\st \g_{\rho\nu}=\g^{\sigma\rho}\g_{\rho\nu}=\delta^\sigma_\nu$ because either the left or the right leg of \eqn{legs} acts  trivially.
Similarly the $\st$-product drops out from the Christoffel symbols \eqn{Chr} and from the curvature expression \eqn{curvcoeff}. The NC Einstein equation reduces then to the commutative one,
and property {\it ii)} follows. 
\end{proof}
\sk

We remark that the NC covariant derivative $\dds_\mu$ differs from the undeformed one $\dd_\mu$. This is due to the different Leibniz rule, 
\eqa
\dds_\mu (h\st \partial_\rho)&=&\partial_\mu h\st\partial_\rho+h\st\Gamma^\st _{\mu\rho}{}^\nu\st\partial_\nu
=\partial_\mu h\;\partial_\rho+h\st\Gamma _{\mu\rho}{}^\nu\,\partial_\nu\label{leibdds}
\\[.6em]
\dd_\mu (h\st\partial_\rho)&=&\dd_\mu (h\partial_\rho)=
\partial_\mu h\;\partial_\rho+h\Gamma _{\mu\rho}{}^\nu\partial_\nu\not= 
\dds_\mu (h\st \partial_\rho)
\ena
The covariant derivative $\dds_\mu$ can consistently be extended along any vector field $u=u^\mu\partial_\mu=u^\mu\st\partial_\mu$ by defining, for all $v\in \Xi$
\eq
\dds_uv=u^\mu\st\dds_\mu v~.\label{dd0}
\en  
This introduces another source of difference, $\dds_uv=u^\mu\st\dds_\mu v\not=
\dd_uv$.
Because of \eqn{leibdds} and of \eqn{dd0}, NC geodesic motion is expected to differ from the undeformed one. 

\section{Geometry and Gravity solutions II: Abelian Twist}
We generalize the results of the previous section to the case of any manifold $M$
with abelian twist, i.e., with a twist $\FF$ of the form
\eq
\FF=e^{-{i\over 2}\theta^{ab}X_a\otimes X_b}~,
\en
where the vector fields $\{X_a\}$ are mutually commuting $[X_a, X_b]=0$. 
This property implies the associativity of the corresponding $\st$-product between functions \eqn{starprodf}, and between tensors \eqn{fhfg3}.

The abelian twist case can be reduced to the previous Moyal-Weyl twist case.  
Let 
\eq
|span\{X_a\}|_{_P}
\en
be the dimension of the vector space spanned by the commuting vector fields $\{X_a\}$ at point $P\in M$.  

We call  $P$ a regular point of the twist $\FF$ if there is an open neighbourhood of $P$ where $|span\{X_a\}|$ is constant.
  Notice that this open neighborhood is itself made of regular points,  and hence the set of all regular points is an open submanifold of $M$. We denote it by $M_{reg}$.

If $P$ is a regular point
we can always write $\theta^{ab}X_a\otimes X_b=\tilde\theta^{\tilde a\tilde b}\tilde X_{\tilde a}\otimes \tilde X_{\tilde b}$ where
the range of the index $\tilde a$ is not greater than that of $a$ and $\tilde X_{\tilde a}$ are linearly independent vectors in each point of the open neighbourhood of $P$ where $|span\{X_a\}|$ is constant.

It follows that locally around $P$ we can consider coordinates $\{x^\mu\}=\{x^{\tilde a},x^i\}$ (where $i=1,...{\rm{dim_{}}}M-|span\{X_a\}|$) such that $\tilde X_{\tilde a}={\partial\over\partial x^{\tilde a}}$ (Frobenius theorem).  In this coordinate system the twist $\FF$ has the canonical Moyal-Weyl structure
${\mathcal F}=\e^{-\frac{\ii}{2}\theta^{\mu\nu}{\partial_\mu}\otimes{\partial_\nu}}$.
We therefore have a Levi-Civita connection defined via its Christoffel symbols, 
and we  can also apply Theorem 1. 
This holds for any regular point $P\in M$.  
These local properties glue together to give the 
corresponding global ones on  $M_{reg}$. 

\sk
\noi {\it Theorem 2.~} Given a manifold $M$ with abelian twist 
$\FF=e^{-{i\over 2}\theta^{ab}X_a\otimes X_b}$, consider the manifold $M_{reg}$ of regular points of $\FF$ and consider a metric $\g$ of $M_{reg}$.

\noi {\it i)} There exists a unique NC Levi-Civita connection on $M_{reg}$ associated with the metric $\g$. 

\noi {\it ii)} If the  Killing Lie algebra $g_K$ of the metric $\g$ has the twist
compatibility property
\eq
\theta^{a b}
{X_a}\otimes\label{legs2}
{X_b}\in \Xi\otimes g_K +g_K\otimes \Xi~
\en
then the NC curvature and  the NC Ricci curvature of the NC Levi-Civita connection are the undeformed ones. 

\noi {\it iii)} If the metric $\g$ is  an Einstein metric 
then it is  also a NC Einstein metric.
\sk
\begin{proof}
{\it i)} The manifold $M_{reg}$ has an atlas of charts where the twist assumes the canonical
Moyal-Weyl form. Let $U$ and $\tilde U$ be the domains of  charts with coordinates $\{x^\mu\}$
and $\{y^\epsilon\}$ respectively. We use the indices $\mu, \nu, \rho$ for the $x$-coordinates in $U$, and the indices $\epsi, \zeta, \kappa$ for the  $y$-coordinates in $\tilde U$. 
We show that if $U\cap \tilde U\not=\varnothing$ then the transition functions $\la_{\epsi}{}^\mu$ defined in $U\cap \tilde U$ by
\eq\label{transfla}
{\partial \over \partial y^\epsi}=\la_{\epsi}{}^\mu\st{\partial\over\partial x^\mu} 
\en
are invariant under the action of the $\st$-product, i.e., for all $h\in A_\st$,
\eq
h\st \la_{\epsi}{}^\mu=
\la_{\epsi}{}^\mu\st h=h \la_{\epsi}{}^\mu~.\label{stinvtransition}
\en
Indeed,
$h\la_\epsi{}^\mu\partial_\mu=h\partial_\epsi=
h\st\partial_\epsi=h\st(\la_\epsi{}^\mu\st\partial_{\mu_{{}_{}}})=
(h\st\la_\epsi{}^\mu)\st\partial_\mu=
(h\st\la_\epsi{}^\mu)\partial_\mu
$, where we used that the twist acts trivially on the partial derivatives of both 
coordinate systems. The equality $\la_{\epsi}{}^\mu\st h=h \la_{\epsi}{}^\mu$
is similarly proven. [Hint:  set $\tau=\partial_\mu$ and $\tau'=h$ in \eqn{fhfg3}.]

We now prove that we have a globally defined connection on $M_{reg}$.
Indeed the connections 
defined in $U$ and in $\tilde U$ by their Chrisitoffel symbols (cf. \eqn{Chr})  
coincide in the intersection $U\cap \tilde U$, for all $u\in\Xi$,
\eq
\dds_u=\tilde \dd^\st_u ~.
\en
Because of \eqn{dd0} this equality is proven if $\dds_\epsi=\tilde \dd^\st_\epsi$
($\epsi=1,...n={\rm dim}M$). This is indeed the case, since
\eqa
\dds_\epsi\partial_\zeta\nn
&=&\la_\epsi{}^\mu\st\dds_\mu(\la_\zeta{}^\nu\st\partial_\nu)\nn\\[.2em]
&=&\la_\epsi{}^\mu\st(\partial_\mu\la_\zeta{}^\nu\st\partial_\nu
  +\la_\zeta{}^\nu\st\Gamma^\st_{\mu\nu}{}^\rho\st\partial_\rho)\nn\\[.2em]
&=&\la_\epsi{}^\mu (\partial_\mu\la_\zeta{}^\nu\, \la_\nu{}^\kappa
  +\la_\zeta{}^\nu\Gamma^\st_{\mu\nu}{}^\rho\la_\rho{}^\kappa)\partial_\kappa\nn\\[.2em]
&=&\tilde\Gamma^\st_{\epsi\zeta}{}^\kappa\partial_\kappa\nn\\[.2em]
&=&\tilde\dd^\st_\epsi\partial_\zeta~
\ena
where in the third line we used property \eqn{stinvtransition}, and
in the fourth line we used that the noncommutative Christoffel
symbols transform under \eqn{transfla} as in the undeformed case. This holds
because the metric and its $\st$-inverse transform as in the undeformed case
(use associativity of the $\st$-product and \eqn{stinvtransition}), and the transition functions $\la_\epsi{}^\mu$, their inverses $\la_\mu{}^\zeta$ and their partial
derivatives are unaffected by the $\st$-product that appears in the NC Christoffel symbols.

\noi {\it ii)} 
The NC curvature and Ricci tensors on $M_{reg}$ of the globally defined NC Levi-Civita connection are the undeformed ones because
equation \eqn{legs2} implies that the hypothesis \eqn{legs} of Theorem 1 holds in an open neighbourhood of any regular point $P$. 

Property {\it iii)} follows immediately from Theorem 1 and the observation that 
the Einstein metric condition \eqn{EinsteinGen} holds in $M_{reg}$ if for any point $P\in M_{reg}$ there exists an open coordinate neighbourhood where the equivalent condition \eqn{Einsteincoeff}  holds.
\end{proof}
\sk
In general $M_{reg}\not=M$, and actually the case $M_{reg}=M$ is a special one
(in the semiclassical limit $M$ is a regular Poisson manifolds).
A typical example where $M_{reg}\not=M$ is provided by the so-called Manin's plane or quantum plane.
A star product that implements the quantum plane commutation relation
$xy=qyx$ ($q=e^{i\hbar}$) can be obtained via the twist 
\begin{equation}
\FF=e^{-{i\over 2}\hbar(x{\partial\over \partial x}\otimes y{\partial\over
\partial y}-y{\partial\over \partial y}\otimes x{\partial\over \partial x})}.
\end{equation}
The vector fields are $x{\partial\over \partial x}$ and
$y{\partial\over \partial y}$. The  irregular points are  the $x$ and $y$ coordinate axes.
Another example is provided by the twists on group manifolds introduced by
\cite{p1Reshetikhin}.
 When the groups are nonabelian then there necessarily exist irregular points of the group manifold.

We now use continuity arguments to extend Theorem 2 from $M_{reg}$ to all $M$. The first step is to show that (as for the regular points of a Poisson structure) the closure of $M_{reg}$ is $M$. The second step is to show that the NC connection has a unique extension from $M_{reg} $ to $M$.
\sk
\noi {\it Theorem 3.}
The open submanifold $M_{reg}$ is dense in $M$.
\begin{proof}
Let $n={\rm dim_{}}M$, and let $R_n$ be the open submanifold of 
regular points $P$ of $M$ such that $|span\{X_a\}|_P=n$.
The condition $|span\{X_a\}|_P=n$  is equivalent to require the matrix $X$ of entries $X_a^\mu$ to have rank $n$.
The entries $X_a^\mu$ are the coefficients of the vector fields $X_a$ with respect to a local frame $\partial_\mu$ (and of course the rank is independent from the choice of the frame).  In turn the rank $n$ condition can be expressed by requiring the sum of 
the squares of all $n^{\rm{th}}$-order minors\footnote{we recall that the $n^{\rm{th}}$-order minors of a given  matrix are the determinants of  the $n\times n$ submatrices of the given matrix.} of the matrix  $X$
to be different from zero. We denote this function by $\sigma_{X,n}$.
In formulae we have 
\eq
|span\{X_a\}|=n ~\Leftrightarrow ~\sigma_{X,n}\not=0~.
\en
Since the vector fields $X_a$ are smooth, 
$\sigma_{X,n} : M\rightarrow \mathbb{R}$ is a smooth function, in particular it is a continuous function and therefore if $\sigma_{X,n}(P)\not=0$ then there is a neighbourhood of $P$ where $\sigma_{X,n}\not=0$. Thus
\eqa
R_n 
&\equiv&\{P\in M ; {\mbox{there exists an open neighbourhood of }} P {\mbox{ where }}  |span\{X_a\}|=n\}\nn\\
&=&\{P\in M ; {\mbox{there exists an open neighbourhood of }} P {\mbox{ where }}  
\sigma_{X,n}\not=0\}\nn\\ 
&=&\{P\in M ; \sigma_{X,n}(P)\not=0\}~.\label{continuity1}
\ena

If $R_n=M$ the theorem is proven. If $R_n\not=M$ we denote by $\overline R_n$ its closure and let $R_{n-1}$ be
the open submanifold of  all regular points $P$
of the complement ${\overline R_n}'_{}$
such that $|span\{X_a\}|_P=n-1$. 
Proceeding as before we have
\eqa
R_{n-1} &\!\!\!\equiv\!\!\!&\{P\in {\overline R_n}'_{} ; {\mbox{there exists an open neighbourhood of }} P {\mbox{ where }}  |span\{X_a\}|=n-1\}\nn\\
&\!\!\!=\!\!\!&\{P\in \overline R_n'{} ; \sigma_{X,{n-1}}(P)\not=0\}~.
\ena
Notice that from the definition of $R_{n-1}$ it follows $R_{n-1}\subset M_{reg}$ because
${\overline R_n}'_{}$ is open in $M$.

 If $R_{n-1}=\overline R_n'$ then
 $R_{n-1}\cup R_n=\overline R_n'\cup R_n$ is dense in $\overline R_n'\cup \overline R_n=M$, and the theorem is proven. 
If $R_{n-1}\not=\overline R_n'$ then
let  ${\overline R_{n-1}}_{}$ be the closure in $\overline R_n'$ of $R_{n-1}$, let 
${\overline R_{n-1}'}_{}$ be its complement in $\overline R_n'$  
and let $R_{n-2}$ be the open submanifold of  all regular points $P$
of ${\overline R_{n-1}'}_{}$  such that $|span\{X_a\}|_P=n-2$.  

If $R_{n-2}=\overline R_{n-1}'$ then
 $R_{n-2}\cup R_{n-1}=\overline R_{n-1}'\cup R_{n-1}$ which is dense in $\overline R_{n-1}'\cup \overline R_{n-1}=\overline R_n'$; therefore
 $R_{n-2}\cup R_{n-1}\cup R_n$ is dense in $\overline R_{n}'\cup R_{n}$ and hence dense in $M$ because $\overline R_{n}'\cup R_{n}$ is dense in $M$ (we use that if $A$ is dense in $B$ and $B$ is open and dense in $C$ then $A$ is dense in $C$).
If $R_{n-2}\not=\overline R_{n-1}'$ then we consider $R_{n-3}\subset \overline R_{n-2}'$ and so on (see Fig. 1).
This iteration procedure stops at the latest when we consider $R_0$. 

If $R_0=\overline R_1'$ the theorem is proven. We now show that
$R_0=\overline R_1'$ always holds. Indeed $R_1$ is the inverse image 
under $\sigma_{X,1}: \overline R_2'\rightarrow \mathbb{R}$ 
of the open interval $\mathbb{R}-\{0\}$, 
\[
R_1\,=\,\{P\in \overline R_2'; \;\sigma_{X,1}(P)\not=0\}\,=\,\sigma_{X,1}^{-1}(\mathbb{R} -\{0\})~.
\]
Hence 
\eqa\overline R_1&=&\{P\in \overline R_2'; \mbox{ for all } I_P,\, I_P\cap \sigma_{X,1}^{-1}(\mathbb{R}-\{0\})\not=\varnothing\}\nn\\[.2em]
&=&
\{P\in \overline R_2'; {\mbox{ for all }} I_P,  \,\sigma_{X,1}(I_P)\not=\{0\}\,\}~,\nn
\ena
where $I_P$ is an open neighbourhood of $P$. It follows that
\eq
\overline R_1'=\{P\in \overline R_2'; {\mbox{ there exists an }} I_P {\mbox{ such that }} \sigma_{X,1}(I_P)=\{0\}\,\}=R_0~.\nn 
\en
\end{proof}
We find practical to summarize the iteration procedure of Theorem 3 in Fig. 1.

\eq
\begin{array}{cccccccccccccc}
& & & & & & & & & & & R_n& \!\!\!\!\subset\!\!\!\! & M  \\[.2em]
& & & & & & & & & R_{n-1}& \!\!\!\!\subset\!\!\!\! & \overline R_n'&  &  \\[.2em]
& & & & & & & R_{n-2}& \!\!\!\!\subset\!\!\!\! & \overline R_{n-1}'&  &  & &\\[.3em]
& & & & & \vdots& ~~& \vdots&  &  & & & &\\[.4em]
& & & R_1& \!\!\!\!\subset \!\!\!\! & \overline R_2'   & & &  &  & & &\\[.2em]
&  R_0& \!\!\!\!=\!\!\!\! & \overline R_1' &  & & & & & &  &  & &
\end{array}\nn
\en
\begin{center}
{\sl Fig. 1.~~The iteration procedure of Theorem 3.}
\end{center}
\sk

In order to extend the connection from $M_{reg}$ to $M$ we recall (see \cite{Teleman})
that the covariant derivative $\dd^\star_u$ defines the connection $\dds$
from vector fields to vector fields with values in 1-forms,\footnote{in the commutative case we should write $\Gamma (TM\otimes T^*M)$ rather than the more intuitive expression $\Om\otimes \Xi$. For smooth second countable finite dimensional manifolds $M$ these two expressions coincide (see for ex. \cite{varilly}, Proposition 2.6). This follows from the existence of a finite covering of $M$ that trivializes the tangent bundle $TM$ and the cotangent bundle $T^*M$, see for example \cite{conlon}, theorem 7.5.16.}
\eqa
\dds : \Xis&\rightarrow& \Om_\st\otimes_\st\Xis\nn\\
             v&\mapsto& \dds(v)\label{dd1}
\ena
where
\eqa
\dds(v) : \Xis&\rightarrow& \Xis\nn\\
             u~&\mapsto& \le u,\dds(v)\res=\dds_u(v)\label{dd2}
\ena
and the nondegenerate pairing $\le~,~\res$ between 1-forms $\om\in \Om_\st$ (or 1-forms with values in vector fields like $\dds(v)$) and vector fields $u\in \Xi_\st$ is given by $\le u,\om\re_\st=\le\of^\al(u),\of_\al(\om)\re$, 
(see Section 5 for a full discussion of the pairing and of equations \eqn{dd1}-\eqn{prop_Delta0}).
The connection $\dds$ satisfies the Leibniz rule, for all functions $h$ and vector fields $v$,
\begin{equation} 
\label{prop_Delta0} 
\dd^\star(h\star v) = \d h \otimes_\star v + h \star
\dd^\star(v)~.
\end{equation}

\sk\noi{\it Theorem 4.} Consider a NC  (pseudo-)Riemannian manifold $M$ with metric $\g$ and abelian twist $\FF=e^{-{i\over 2}\theta^{ab}X_a\otimes X_b}$.
Consider also the associated NC  (pseudo-)Riemannian manifold $M_{reg}$ and the
NC Levi-Civita connection $\dds$ on $M_{reg}$ uniquely  defined by the Christoffel symbols \eqn{Chr}. This connection has a unique smooth extension $\hat\dd^\st$ to $M$. This extension is the NC Levi-Civita connection on $M$.
\sk
\begin{proof}
In Section 5.5, Theorem 5, we show that the torsion free and metric compatibility conditions uniquely determine the expression of $\hat \dd^\st$ on two arbitrary vector fields
$u$ and $v$. This is the case at zeroth order in the noncommutativity parameter $\theta$, and the higher order $\theta$-terms in $\hat\dd^\st$ can be 
perturbatively calculated. This shows the unicity of the connection $\hat\dd^\st$.
It also shows that $\hat\dd^\st_u(v)$ is a smooth vector field if $u$ and $v$ are smooth,
and that $\hat\dd^\st$ has local properties, i.e.
if $u=u'$ and $v=v'$ in an open $U\in M$ then 
\eq
\hat\dd^\st_u(v)=\hat\dd^\st_{u'}(v')
\label{local}
\en
 in $U\in M$. In order to prove that this map $\hat \dd^\st$ is a connection we have to show that 
it satisfies the Leibniz rule. We proceed in three steps.
 
\noi 1)
It is easy to see that the covariant derivative $\dds_\mu$ on $M_{reg}$, uniquely  defined by the Christoffel symbols \eqn{Chr},
defines a connection $\dds$ on $M_{reg}$; indeed the Leibnitz rule for $\dds$
immediately follows from the $\dds_\mu$ Leibniz rule \eqn{leibdds}. 

\noi
2) In $M_{reg}$ both $\hat\dd^\st$ and $\dds$ are metric compatible and torsion free
and therefore they coincide. 

\noi 3) The unique smooth extension $\hat\dd^\st$ of $\dds$ is a connection
because the Leibnitz rule 
\eq
\hat\dd^\star(h\star v) = \d h \otimes_\star v + h \star
\hat \dd^\star(v)~
\en
holds  for any point $P\in M_{reg}$ and hence by continuity for any point  $P\in M$.
\end{proof}

\sk
\noi {\it Remark.~~}Theorem 4 is the fundamental theorem of NC Riemannian Geometry with noncommutativity given by an arbitrary abelian twist  $\FF=e^{-{i\over 2}\theta^{ab}X_a\otimes X_b}$.
\sk
Continuity of the NC Riemann, Ricci and metric tensors leads immediately to generalize to $M$ the results found for $M_{reg}$ in Theorem 2;
\sk
\noi {\it Corollary 1.~}{{Given a NC manifold $M$ with abelian twist 
${\mathcal F}=\e^{-\frac{\ii}{2}\theta^{ab}
{X_a}\otimes {X_b}}$,  consider all metrics $\g$  such that the associated Killing Lie algebra $g_K$ has the twist compatibility property}},
\eq 
\theta^{ab}
{X_a}\otimes\nonumber
{X_b}\in \Xi\otimes g_K +g_K\otimes \Xi.
\en
\noi
{\it i)} The NC Riemann tensors and the NC Ricci tensors of the NC
Levi-Civita connections of these metrics $\g$ are the undeformed ones.

\noi {\it ii)} If these metrics $\g$ are Einstein metrics then they are also NC Einstein
metrics.~~~~~~~
{\small{$\square$}}

\section{Drinfeld Twist Differential Geometry}
In this section we consider a general (triangular) Drinfeld twist on a maniolfd $M$
and construct the corresponding noncommutative differential and Riemannian geometry.
The results obtained, in particular those regarding the properties of  the contraction operator and pairing between covariant and contravariant tensors, and the uniqueness and smoothness of the NC Levi Civita connection, are needed in order to construct explicit  NC differential geometry  solutions.

\subsection{Deformation by twists}
We begin by recalling the general setting 
used to introduce a star product via a twist. 

Consider a Lie algebra $g$ over $\cc$, and its associated universal enveloping algebra $Ug$.  We recall that the elements of $Ug$ are 
the complex numbers $\cc$ and sums 
of products of elements $t\in g$, where we identify 
$tt'-t't$ with the Lie algebra element $[t,t']$. 
$Ug$ is an associative algebra with unit. It is a Hopf algebra with coproduct 
$ \Delta:Ug\rightarrow Ug\otimes Ug~$, counit $\epsi : Ug\rightarrow \cc$
and antipode $S$ given on the generators as:
\begin{eqnarray}
 & & \D(t)=t \otimes 1 + 1 \otimes t~~~\D(1)=1\otimes 1  
  \label{copL}\\
 & & \epsi (t)=0~~~~~~~~~~~~~~~~~~~~\epsi (1)=1 \label{couL}\\
 & & S(t)=-t~~~~~~~~~~~~~~~~~S (1)=1 \label{coiL}
\end{eqnarray}
and extended to all $U(g)$ by requiring 
$\D$ and $\epsi$ to be linear and multiplicative
(e.g. $\D(tt'):=\D(t)\D(t')=tt'\otimes 1+t\otimes t' +t'\otimes t+ 1\otimes tt'$),
while $S$ is linear and antimultiplicative.
In the sequel we use Sweedler coproduct notation
\eq
\Delta(\xi)=\xi_1\otimes \xi_2
\en
where $\xi\in Ug$,  $\xi_1\otimes \xi_2\in Ug\otimes Ug$ and a sum
over $\xi_1$ and $\xi_2$  is understood.

We extend the notion of enveloping algebra to formal power series in 
$\la$ (we replace the field $\cc$ with the ring $\cc[[\la]]$)  and we 
correspondingly consider the Hopf algebra $(Ug[[\lambda]],\cdot,\Delta, S,\varepsi)$.
In the sequel for sake of brevity we denote 
$Ug[[\la]]$ by $Ug$.
\sk
A twist $\FF$ is an element 
$\FF\in Ug\otimes Ug$ that is invertible and that satisfies
\eq\label{propF1}
\FF_{12}(\Delta\otimes id)\FF=\FF_{23}(id\otimes \Delta)\FF\,,
\en
\eq\label{propF2}
(\varepsi\otimes id)\FF=1=(id\otimes \varepsi)\FF~,
\en
where $\FF_{12}=\FF\otimes 1$ and $\FF_{23}=1\otimes \FF$.

We in addition require\footnote{ Actually it is
possible to show that (\ref{consF}) is a consequence  of
(\ref{propF1}), (\ref{propF2}) and of $\FF$ being at each order in
$\la$ a finite sum of finite products of Lie algebra elements} 
\eq
\label{consF}
\FF=1\otimes 1 + {\cal O}(\la)~.
\en
Property (\ref{propF1}) states that $\FF$ is a two cocycle, this property is responsible for the associativity of the $\st$-products to be defined. Property 
(\ref{propF2}) is just a
normalization condition. From (\ref{consF}) it follows that 
${\cal F}$ can be formally inverted as a power series in $\lambda$. It also shows that the geometry we are going to construct has the nature of a deformation, i.e. in the $0$-th order in $\lambda$ we recover  the usual undeformed geometry.

In the notation (sum over $\al$ understood)
\eq\label{Fff}
\FF=\f^\al\otimes\f_\al~~~,~~~~\FF^{-1}=\of^\al\otimes\of_\al~,
\en
the elements $ \f^\alpha, \f_\alpha, \of^\al,\of_\al$ belong to  $Ug$. 

In order to become more familiar with this notation we  rewrite equation
(\ref{propF1}) and its inverse, 
\eq\label{ifpppop}
((\Delta\otimes id)\FF^{-1}) 
\FF^{-1}_{12} =((id \otimes \Delta)\FF^{-1})\FF^{-1}_{23}~,
\en 
as well as (\ref{propF2}) and (\ref{consF}) using the notation (\ref{Fff}):
\begin{eqnarray}
\f^\beta \f^\alpha_{_1}\otimes \f_\beta \f^\alpha_{_2}\otimes \f_\alpha &=& 
\f^\alpha\otimes \f^\beta \f_{\alpha_1}\otimes \f_\beta
\f_{\alpha_2} ,\label{2.21}\\
\label{ass}
\of_{_1}^\al\of^\be\otimes \of_{_2}^\al\of_\be\otimes \of_\al&=&
\of^\al\otimes {\of_{\al_1}}\of^\be\otimes {\of_{\al_2}}\of_\be~,\\
\varepsilon(\f^\alpha)\f_\alpha &=& 1 =  \f^\alpha\varepsilon(\f_\alpha) , \label{2.23}\\
{\cal F} = \f^\alpha\otimes \f_\alpha &=& 1\otimes 1 + {\cal O}(\lambda) .\label{2.24}
\end{eqnarray}

\sk
Consider now an algebra $A$ (over $\cc[[\la]$), 
and an action of the Lie algebra $g$ on $A$, $a\mapsto t(a)$ where $t\in g$ 
and $a\in A$. We require compatibility of this action with the product in $A$
i.e.  for any $t\in g$ we have a derivation of $A$, 
\eq
t(ab)=t(a)b+at(b)~.\label{compprim}
\en
The action of $g$ on $A$ induces an action of the universal enveloping 
algebra $Ug$ on $A$  (for example the element $tt'\in Ug$ has action 
$t(t'(a))$). We say that $A$ is a $Ug$-module algebra, i.e.,
the  algebra structure of the $Ug$-module $A$ 
is compatible with the $Ug$ action,
for all $\xi\in \UU$ and $ a,b\in A$,
\eq\xi(ab)=\mu\circ\Delta(\xi)(a\otimes b)=\xi_1(a)\xi_2(b)~~~,~~~~
\,\xi(1)=\varepsilon(\xi)1\,. \label{Ugmodulealg}
\en
(where $1$ is the unit in $A$). This property is equivalent to  
\eqn{compprim}. 
\sk
Given a twist $\FF\in Ug\otimes Ug$,
we can construct a deformed algebra $A_\st$.
The algebra $A_\st$ has the same vector space 
structure as $A$. 
The product in $A_\st $ is defined by
\eq
a\st  b=\mu\circ \FF^{-1}(a\otimes b)=\of^\al(a)\of_\al(b)~.
\en
In order to prove associativity of the new product we
use (\ref{ass}) and compute:
\eqa
(a\st b)\st c &=&\of^\al(\of^\be(a)\of_\be(b))\of_\al(c)=
(\of_{_1}^\al\of^\be)(a) (\of_{_2}^\al\of_\be)(b) \of_\al(c)\nn
=
\of^\al(a) ({\of_{\al_1}}\of^\be)(b)
({\of_{\al_2}}\of_\be)(c)\nn \\
&=&
\of^\al(a)\of_\al(\of^\be(b)\of_\be(c))=a\st (b\st c)\nn~.
\ena

Examples of twists include the Moyal-Weyl and the abelian twists of the previous sections.  Twists however are not necessarily related to abelian Lie algebras. 
For example consider the elements $H, E,A,B$, that satisfy the Lie algebra relations
\eqa
& & [H,E]=2E~,~~ [H,A]=\al A~,~~[H,B]=\be B~,~~~~~~~~\al+\be=2~,\nn\\[.1cm]
& & [A,B]=E~,~~~~[E,A]=0~,~~~\;~[E,B]=0~.   
\ena
Then the element
\eq
\FF=e^{{1\over 2}H\otimes_{} {\rm{ln}}(1+\la E)} \:  e^{\la A\otimes B{1\over 1+\la E}}
\en
is a twist and gives a well defined $\st$-product on the algebra of
functions on $M$. When $\la\rightarrow 0$ we recover the undeformed product.  
These twists are known as extended Jordanian deformations
\cite{Kulish1}. Jordanian deformations 
\cite{GER, OGIEV} are obtained setting $A=B=0$ (and keeping the relation 
$[H,E]=2E$).

\subsection{$\st$-Noncommutative Manifolds}

\noi  A $\st$-noncommutative manifold is a smooth manifold $M$ 
with a twist $\FF\in U\Xi\otimes U\Xi$, where $\Xi$ is the Lie algebra of vector fields on $M$.

We now use the twist to  deform the commutative geometry on a manifold $M$
(vector fields, 1-forms, exterior algebra, tensor algebra, symmetry algebras, 
covariant derivatives etc.) into the twisted noncommutative one. 
The guiding principle is the observation that every time we have 
a bilinear map $$\mu\,: X\times Y\rightarrow Z~~~~~~~~~~~~~~~~~~$$
where $X,Y,Z$ are vector spaces, and where there is an action of the Lie algebra $g$   
(and therefore of $\FF^{-1}$) on $X$ and $Y$
we can compose this map with the action of the twist. In this way
we obtain a deformed version $\mu_\st$ of the initial bilinear map $\mu$:
\eqa
\mu_\st:=\mu\circ \FF^{-1}~,\label{generalpres}&~~~~~~~~~~~~~~&
\ena
{\vskip -.8cm}
\eqa
{}~~~~~~~~~~~~~\mu_\st\,:X\times  Y&\rightarrow& Z\nn\\
(\x, \y)\,\, &\mapsto& \mu_\st(\x,\y)=\mu(\of^\al(\x),\of_\al(\y))\nn~.
\ena
The action of the Lie algbra $\Xi$ of vector fields  on the vector spaces $X,Y,Z$ we consider will always be via the Lie derivative.

Using \eqn{generalpres}, noncommutative functions $A_\st$, vector fields $\Xi_\st$  and  tensor fields $\TT_\st$ have been defined in Section 2.

We  next introduce the universal
$\RR$-matrix \eq \RR:=\FF_{21}\FF^{-1}~\label{defUR} \en where by
definition $\FF_{21}=\f_\al\otimes \f^\al$. In the sequel we use
the notation \eq
\RR=\R^\al\otimes\R_\al~~~,~~~~~~\RR^{-1}=\oR^\al\otimes\oR_\al~.
\en  The $\RR$-matrix measures
the noncommutativity of the $\star$-product. Indeed it is easy to
see that \eq\label{Rpermutation} h\st g=\oR^\al(g)\st\oR_\al(h)~.
\en The permutation group in noncommutative space is naturally
represented by $\RR$. Formula (\ref{Rpermutation}) says that the
$\st$-product is $\RR$-commutative .
 \sk
\sk
\noi{\bf Exterior forms 
$\Omega^{\mbox{\boldmath $\cdot$}}_\st=\oplus_p\Omega^{p}_\st$}.
Exterior forms form an algebra with product 
$\wedge :\,\Omega^{\mbox{\boldmath $\cdot$}}\times 
\Omega^{\mbox{\boldmath $\cdot$}}\rightarrow \Omega^{\mbox{\boldmath $\cdot$}}$.
We $\st$-deform the wedge product into the $\st$-wedge product,
\eq\label{formsfromthm}
\vartheta\wedge_\st\vartheta':=\of^\al(\vartheta)\wedge \of_\al(\vartheta')~.
\en 
We denote by $\Omega^{\mbox{\boldmath $\cdot$}}_\st$ 
the linear space of forms equipped with the wedge product  
$\wedge_\st$.

As in the commutative case exterior forms are totally 
$\st$-antisymmetric contravariant tensor fields. 
For example the 2-form $\omega\wedge_\st\omega'$ is 
the $\st$-antisymmetric combination
\eq\label{stantisymm}
\omega\wedge_\st\omega'= \omega\otimes_\st\omega'
-\oR^\al(\omega')\otimes_\st \oR_{\al}(\omega)~.
\en
The exterior derivative 
${\rm d}:A\rightarrow \Om$ satisfies the Leibniz rule 
${\rm d}(h\st g)={\rm d} h\st g+h\st{\rm d}g $, indeed 
\eqa 
{\rm d}(h\st g)={\rm d} (\of^\al (h) \,\of_\al( g))&=&
{\rm d} (\of^\al (h))\, \of_\al (g)+\of^\al( h)\, \d\of_\al( g)\nn\\
&=&
\of^\al({\rm d} h)\, \of_\al (g)+\of^\al( h)\, \of_\al (\d g)\nn\\
&=&{\rm d} h\st g+h\st{\rm d}g \label{dhgLeib}
\ena
where we observed that for each index $\al$, $\of^\al$ and $\of_\al$ are products 
of vector fields acting via the Lie derivative on functions or 1-forms, and that
therefore commute with the exterior differential because 
$\LL_{uv\ldots z}\equiv\LL_u\circ \LL_v\circ \ldots\LL_z$ and the Lie derivative along a vector field commutes with the differential.

The usual exterior derivative is therefore also the $\st $-exterior derivative. 
\sk
\noi{\bf $\st$-Pairing}. 
We now consider the bilinear map 
$
\langle ~,~\re : \,
\Xi\times \Omega \rightarrow  A\,,$
$(v,\omega)~\mapsto \langle  v,\omega\re$,
where, using local coordinates,
$\langle  v^\mu\partial_\mu,\omega_\nu dx^\nu\re=v^\mu\om_\mu~.$
Always according to the general prescription (\ref{generalpres}) we deform 
this pairing into
\eqa\label{lerest}
\langle ~,~\re_\st : \,
\Xi_\st\times \Omega_\st &\rightarrow & A_\st~,\\
(\xi,\omega)~&\mapsto &\langle  \xi,\omega\re_\st
:=\langle \of^\al(\xi),\of_\al(\omega)\re~.
\ena
It is easy to see that due to the cocycle condition for $\FF$ 
the $\st$-pairing satisfies the  
$A_\st$-linearity properties  
\eq
\langle  h\st u,\omega\st k\re_\st=h\st\langle  u,\omega\re_\st\st k~,
\en
\eq\label{linearp}
\langle  u, h\st\omega \re_\st=
{\oR^\al}(h)\st\langle  {\oR_{\al}}(u),\omega \re_\st~.
\en
Using the pairing $\langle ~\,,~\,\res$ we associate with any $1$-form
$\om$ the left $A_\st$-linear map $\langle ~\,,\om\res$. 
Also the converse holds: any left $A_\st$-linear map 
$\Phi:\Xis\rightarrow \AAs$ is of the form $\langle ~\,,\om\res$
for some $\omega$.

\sk
The pairing can be extended to covariant tensors 
and contravariant ones. We first define in the undeformed
case the pairing 
($u^i$ denote vector fields, $\theta^j$ denote $1$-forms),
\eq
\langle u^p\ldots \otimes u^{2}\otimes u^1\,,\,
\theta^1\otimes\theta^2\ldots\otimes \theta^p\rangle =
\langle u^1,\theta^1\rangle \,\langle u^2, \theta^2\rangle\,\ldots\langle u^p, \theta^p\rangle
\en
and more in general the pairing
\eq
\langle u^p\ldots\otimes u^1\,,\,
\theta^1\otimes\ldots \theta^p\otimes \tau\re=\langle u^1,\theta^1\rangle \,\ldots\langle u^p, \theta^p\rangle\,\tau
\en
that is obtained 
by first contracting the innermost elements; here $\tau$ is an arbitrary tensor field. Using locality and  linearity
this pairing is extended to any $p$-covariant tensor $\nu\in\TT^{0,p}$ and  
any tensor $\rho\in\TT^{q,s}$ at least  $p$-times contravariant ($q\geq p$).
It is this onion-like structure pairing that naturally generalizes to the
noncommutative case.

The $\st$-pairing is defined by
\eq
\langle \nu, \rho \rangle_\st :=
\langle \bar\ff^\al(\nu) , \bar\ff_\al(\rho)\rangle~.
\en
Using the cocycle condition for the twist $\FF$ and the onion like structure of the undeformed pairing we have the property
\eq\label{nestedpairing}
\langle \nu\otimes_\st u\,,\, \rho \rangle_\st =
\langle \nu \,,\,\langle u,\rho\rangle_\st\res~.
\en

\sk
\noi {\bf $\st$-Lie algebra of vector fields $\Xi_\st$}. 
The Lie algebra of vector fields is the Lie algebra of 
infinitesimal transformations (infinitesimal local diffeomorphisms). 
Vector fields act on tensor fields via the Lie derivative. The relativity principle of general covariance is implemented as general covariance under infinitesimal diffeomorphisms. These are given by  Lie derivatives.
In the noncommutative case  the Lie algebra of vector fields is deformed, 
and applying the recipe (\ref{generalpres}) we obtain the $\st$-Lie bracket 
\begin{eqnarray}
[\quad ]_\st: \quad\quad \Xi\times\Xi &\to& \Xi \nonumber\\
(u,v) &\mapsto& [u,v]_\st:=[\of^\al(u),\of_\al(v)]~ .\label{2.1st}
\end{eqnarray}
This can be realized as a deformed commutator
\eqa 
[u,v]_\st&=&[\of^\al(u),\of_\al(v)]=\of^\al(u)\of_\al(v)-\of_\al(v)\of^\al(u)
\nn\\
 &=&u\st v-\oR^\al(v)\st\oR_\al(u)~,
\ena
where the $\st$-product between vector fields is given by $u\st v=\of^\al(u)\of_\al(v)$.

It is easy to see that the bracket  $[~,~]_\st $ 
has the $\st$-antisymmetry property
\eq\label{sigmaantysymme}
[u,v]_\st =-[\oR^\al(v), \oR_\al(u)]_\st~ .
\en
This can be shown as follows: 
$[u,v]_\st =[\of^\al(u),\of_\al(v)]=-[\of_\al(v),
\of^\al(u)]=
-[\oR^\al(v), \oR_\al(u)]_\st~. $
A $\st$-Jacoby identity can be proven as well
\eq
[u ,[v,z]_\st ]_\st =[[u,v]_\st ,z]_\st  
+ [\oR^\al(v), [\oR_\al(u),z]_\st ]_\st ~.
\en
 
 In the commutative case the commutator $[u,v]$ equals the Lie derivative $\LL_u(v)$.
It is then natural to define the $\st$-Lie derivative  as
\eq
\ll^\st_u:=\ll_{\of^\al(u)}\circ \of^\al~,\label{defstLieder}
\en 
so that $\LL_u^\st(v)=[u,v]_\st$. Notice that the $\st$-Lie derivative
 is given by combining the usual Lie derivative
with the twist $\FF$ as in (\ref{generalpres}).

Definition \eqn{defstLieder} holds more in general when the $\st$-Lie derivative acts on tensor fields $\TT_\st$.
The $\st$-Lie derivative satisfies the deformed Leibniz rule, for all $\tau,\tau'\in\TT_\st$,
\eq\label{Leiblie}
\ll^\st_u(\tau\otimes_\st \tau')=\ll_u^\st(\tau)\otimes_\st \tau' + 
\oR^\al(\tau)\otimes_\st \ll_{\oR_\al(u)}^\st(\tau')~;
\en
in particular on functions $h,g$ we have
$\,\ll^\st_u(h\st g)=\ll_u^\st(h)\st g + 
\oR^\al(h)\st \ll_{\oR_\al(u)}^\st(g)~.
$

\subsection{Covariant Derivative}\label{covderivative}

A connection is a linear mapping 
\begin{equation} \dd^\star : \Xi_\star \rightarrow
\Omega_\star \otimes_\star \Xi_\star 
\end{equation} 
which satisfies the (undeformed) Leibniz rule, for all $v\in\Xis$, 
\begin{equation} 
\label{prop_Delta} 
\dd^\star(h\star v) = \d h \otimes_\star v + h \star
\dd^\star(v)~.
\end{equation}

Associated with a connection $\dd^\star$ we have the covariant
derivative 
$\dd^\star_u$ along the vector field $u\in \Xi_\st$. It is defined by, for
all $v\in\Xis$ 
\begin{equation}\label{covariant_derivative} 
\dd^\star_u(v) := \langle
  u,\dd^\star v \rangle_\st ~.
\end{equation} 
From (\ref{prop_Delta}) and (\ref{covariant_derivative}) we
immediately have, for all $u,v,z\in\Xi_\st,~ h\in A_\st, 
$:
\eqa
&&\dd_{u+v}^{\star}z=\dd_{u}^{\star}z+\dd_{v}^{\star}z~,\\[.35cm]
&&\dd_{h\star u}^{\star}v=h\star\dd_{u}^{\star}v~,\\[.35cm]
&&\dd_{u}^{\star}(v+z)=\dds_{u}(v)+\dd_{u}^{\star}(z)~,\\[.35cm]
&&\dd_{u}^{\star}(h\star v)
\,=\,\mathcal{L}_u^{\star}(h)\star v+ \oR^\al(h)\st\dd^\st_{\oR_\al(u)}v~\label{covlast}.
\ena
We notice that the covariant derivative $\dd^\st_u$ satisfies the same deformed
Leibniz rule as the Lie derivative $\mathcal{L}^\st_u$ (cf. \eqn{Leiblie}). As in the undeformed case we define the covariant derivative on functions to be equal to the Lie derivative, for all $h\in A_\st$,
\eq
\dds_u(h)=\lls_u(h)~.
\en 
We also notice that the covariant derivative is defined only along vector fields and not along products of vector fields. The right hand side of expression \eqn {covlast} is well defined because of the peculiar property of the
Leibniz rule \eqn{Leiblie}: 
$\oR_\al(u)$ is again a vector field. 
\sk
With respect to a local frame of vector fields 
$\{e_i\}$ 
we have the connection coefficients
\eq\label{Connectioncoeffijk}
\dds_{e_i} e_j={\Gamma_{ij}}^k\st e_k~.
\en

\noi{\bf Covariant derivative on tensor fields.}  We define the covariant derivative on covariant tensors by iterated use of
the following deformed Leibniz rule \cite{AS}, for all $u,v, z\in \Xis$,
\eq \label{Leibdds}
\dds_u(v\otimes_\st z):= \oR^\al(\dds_{\oR_\be(u)}\oR_\gamma(v))\ots \oR_\al\oR^\be\oR^\gamma (z) + \oR^\al(v)\ots 
\dds_{\oR_\al(u)}z~.
\en
As in the commutative case we define the covariant derivative on 1-forms $\Oms$ by requiring compatibility with the contraction operator, for all $u,v\in\Xis, \omega\in\Omega_\st$, 
\eq\label{compleredd}
\dds_u\le v,\om\res =\le\oR^\al(\dds_{\oR_\be(u)}\oR_\gamma(v)), \oR_\al\oR^\be\oR^\gamma (\om)\res + 
\le\oR^\al(v),\dds_{\oR_\al(u)} \om\res
\en
so that
$
\le v,\dds_u \om\res=
\lls_{\oR^\al(u)}\le \oR_\al(v),\om\res
-\le\oR^\al(\dds_{\oR_\be\oR^\de(u)}\oR_\gamma\oR_\de(v)), \oR_\al\oR^\be\oR^\gamma (\om)\res 
\,.  
$
Finally we extend the covariant derivative to all tensor fields via 
the deformed Leibniz rule (\ref{Leibdds}) where now 
$\tau, \tau' \in \TT_\st$,
\eq\label{Leibddsg}
\dds_u(\tau\ots\tau'):=\oR^\al(\dds_{\oR_\be(u)}\oR_\gamma(\tau))\ots \oR_\al\oR^\be\oR^\gamma (\tau') + \oR^\al(\tau)\ots 
\dds_{\oR_\al(u)}\tau'~.
\en

\subsection{Torsion, Curvature and Ricci tensor}
The torsion $\tr$ and the curvature $\rr$ associated with
a connection $\dd^\st$ are 
defined by
\eqa
\tr(u,v)&:=&\dd_{u}^{\star}v-\dd_{\oR^{\alpha}(v)}^{\star}\oR_{\alpha}(u)
-[u,v]_{\star}~,\\[.2cm]
\rr(u,v,z)&:=&\dd_{u}^{\star}\dd_{v}^{\star}z-
\dd_{\oR^\al{(v)}}^{\star}\dd_{\oR_\al(u)}^{\star}z-\dds_{[u,v]_\st} z~,
\ena
for all $u,v,z\in\Xis$.

The presence of the $R$-matrix in the definition of torsion and curvature ensures 
that  $\tr$ and $\rr$  are left $A_\st$-linear maps, i.e.
$$
\tr(f\star u,v)=f\star \tr(u,v)~
~~,~~~\tr(u,f\star v)= \oR^\al (f)\st\tr(\oR_\al(u) ,v)
$$
and similarly for the curvature.
The $A_\st$-linearity of $\tr$ and $\rr$ ensures that we
have a well defined  torsion tensor and  curvature  tensor.  We denote them by the same letters $\tr$ and $\rr$; they are given by, for all $u,v\in \Xi_\st$, $\om\in \Om_\st$,
\eqa
&&\le u\otimes_\st v \,, \tr \res = \tr(u,v)~,\\
&&\le u\otimes_\st v\otimes_\st z\, ,\rr\res=\rr(u,v,z)~.
\ena

\noi{\bf Local coordinates description.} We denote by $\{e_i\}$ a local frame of vector fields 
(subordinate to an open $U\subset M$)
and by $\{\theta^j\}$ the dual frame of 1-forms:
\eq
\langle  e_i\,,\,\theta^j\re_\st=\delta^j_i~.
\en
The coefficients ${\tr_{ij}}^l$ and ${\rr_{ijk}}^l$ of the torsion and curvature tensors
with respect to this local frame are uniquely 
defined by the following expressions
\eqa
&&\tr=\theta^j\otimes_\st\theta^i\st{\tr_{ij}}^l\otimes_\st e_l\label{Tijk}~,\\
&&\rr=\theta^k\otimes_\st\theta^j\otimes_\st\theta^i
\st{\rr_{ijk}}^l\otimes_\st e_l\label{Rijkl}~,
\ena
so that
$
{\tr_{ij}}^l=\langle \tr(e_i,e_j)\,,\,\theta^l\re_\st~,~~
{\rr_{ijk}}^l=\langle \rr(e_i,e_j,e_k)\,,\,\theta^l\re_\st~.
$
We also have \cite{Teleman}
\eqa
&&\tr={1\over 2}_{^{}}\theta^j\wedge_\st\theta^i\st{\tr_{ij}}^l\otimes_\st e_l\label{TTijk}~,\\[.1cm]
&&\rr={1\over 2}_{^{}}\theta^k\otimes_\st\theta^j\wedge_\st\theta^i
\st{\rr_{ijk}}^l\otimes_\st e_l\label{RRijkl}~.
\ena

\sk
We recall that the  Ricci tensor is given by the following contraction of the
curvature:
\eq\label{defofric1}
\ric(u,v):=\le \theta^i, \rr(e_i,u,v)\res~,
\en
where sum over $i$ is understood. The contraction
$\le~,~\res$ in \eqn{defofric1} is 
a contraction between forms on the {\sl left} and
vector fields on the {\sl right}. It is defined through the  
deformation of the commutative pairing, 
$
\le \om\,,\,u\res=\le\of^\al(\om)\,,\,\of_\al (u)\re~.
$

Definition (\ref{defofric1}) is well given because
it is independent from the choice of the frame $\{e_i\}$ 
(and the dual frame $\{\theta^i\}$),
and because the Ricci map  so defined is an $A_\st$-linear map.

The coefficients of the Ricci tensor are 
\eq
\ric_{jk}=\ric(e_j,e_k)~.
\en
In the commutative limit these tensors become the usual torsion, curvature and Ricci tensors, 
\eq
\tr\rightarrow\trc~,~~\rr\rightarrow \rrc~,~~\ric\rightarrow \ricc~.
\en
and in particular we recover the usual components relation $\ricc_{jk}=\rrc_{ijk}{}^k$.

\subsection{$\st$-Riemannian geometry}
Along these lines one can also consider $\st$-Riemaniann geometry.
In order to define a $\st$-metric we need to define $\st$-symmetric 
elements in $\Oms\ots\Oms$ where $\Oms$ is the space of 1-forms. 
Recalling that permutations are implemented with the $R$-matrix 
we see that $\st$-symmetric elements are of the form
\eq\label{omompr}
\omega\otimes_\st\omega'
+\oR^\al(\omega')\otimes_\st \oR_{\al}(\omega)~.
\en
In particular any symmetric tensor in
$\Om\otimes\Om\,$ is also a $\st$-symmetric tensor in 
$\Oms\ots\Oms$, indeed expansion of \eqn{omompr} gives 
$\of^\al(\om)\otimes\of_\al(\om')+\of_\al(\om')\otimes\of^\al(\om)$
that is a sum (over $\al$) of symmetric tensors.
Similarly for antisymmetric tensors.

In particular, since a commutative metric is a nondegenerate symmetric tensor in
$\Om\otimes\Om\,$
we conclude that any commutative metric is also a noncommutative metric,
($\st$-nondegeneracy of the metric is insured by the fact that at
zeroth order in the deformation parameter 
the metric is nondegenerate). We denote by $\g$ the metric tensor.
If we write 
\eq\g=\g^a\ots\g_a\in\Oms\ots\Oms~\en 
(for example locally $\g=\theta^j\otimes_\st\theta^i\st\g_{ij}$),
then for every $v\in\Xis$ we can define the 1-form
\eq\label{defsubsid}
\le v , \g\res:=\le v,\g^a\res\st\g_a
\en
and we can then construct the left $A_\st$-linear map
$\g$, corresponding to the metric tensor $\g\in\Oms\ots\Oms$, as
\eqa\label{metricoperat}
\g\,:\,\Xis\ots\Xis &\rightarrow&\AAs\nn\\
(u,v)\,\,&\mapsto&\g(u,v)=\le u\ots v,\g\res:=
\le u\,,\le v,\g\res\res~.
\ena

The $\st$-inverse metric
$\g^{-1}\in\Xis\otimes_\st\Xis$ has been defined in \cite{GR1} by first considering the metric as a map from vector fields to 1-forms and then by defining the inverse metric as the inverse of this map.  
If we write $\g=\g^a\otimes_\st \g_a$ and 
\eq\g^{-1}={\g^{-1}}^b\ots\g^{-1}_b\in\Xis\ots\Xis~\en
then 
$\g^{-1}$ is defined by the condition, for all $\om\in\Om_\st$,
\eqa\label{thedefofmetricstinv}
\le\le \om_{},_{}\g^{-1}\res,\,\g\res=\om~.\label{thedefofmetricstinv2}
\ena

We now consider the connection that has vanishing torsion and that
is metric compatible, for all $u\in \Xi_\st$, 
\eq\label{metricdd}
\dds_u\g=0~;
\en
equivalently (recall \eqn{Leibdds}) for all $u,v,z\in\Xis$,
\eq\label{metricddexpl}
\dds_u\,\g(v,z)=\g(\oR^\al(\dds_{\oR_\be(u)}\oR_\gamma(v)), \oR_\al\oR^\be\oR^\gamma (z))+\g(\oR^\al(v),\dds_{\oR_\al(u)} z)~.
\en
For the $\theta$-constant case the explicit expression of the connection is via its Christoffel symbols \eqn{Chr}, cf. \cite{GR1}.  For the case the twist is compatible with the metric as in \eqn{comptwist}, the existence of the Levi-Civita connection $\dds$ is proven in Theorem 8. For the abelian twist case existence 
of the Levi-Civita connection $\dds$ is proven in  Theorem 4 (that uses Theorem 5 below).

We can then define the scalar curvature $\mathfrak{R}^\st$ with respect to this connection. It is given by 
\eq
\mathfrak{R}^\st:=\le \g^{-1},\ric\res~.
\en
Locally we
write $\g^{-1}=\g^{ij_{^{^\st}}}\st e_j\ots e_i$, and 
\eq
\mathfrak{R}^\st=
\g^{ij_{^{^\st}}}\st \ric_{ji}~.
\en
The Einstein tensor is then defined by 
\eq
\G^\st:=\ric-{1\over 2}{_{_{}}}\g\st\mathfrak{R}^\st~,
\en
and Einstein equations in vacuum are 
\eq\label{Einstein}
\G^\st=0~,
\en
or equivalently,\footnote{$\G^\st=0$ implies $\le\g^{-1},\G^\st\res=0$, and $1-{1\over 2}\le\g^{-1},\g\res\not=0$ as is easily seen in the commutative limit.} 
$
\,\,\ric=0~.
$

We also define a $\st$-Einstein manifold to be a $\st$-Riemannian manifold that
satisfies the condition
\eq
\ric=c\,\g
\en
for some real constant $c$ (equivalently we can require the Einstein tensor $\G^\st$ to be proportional to the metric tensor $\g$).

\sk
We conclude this section by showing that if a NC Levi-Civita connection exists,
then it is unique and can be determined by a perturbative expansion order by order in the noncommutativity parameter. If we are able to prove that this unique expression thus obtained satisfies the Leibniz rule $\dds (h\st v)= d h\otimes_\st v+ h\st\dds v$ then
(as for example we do with the hypotheses of  Theorem 4)  we have existence and uniqueness of the NC Levi-Civita connection.

\sk
\noi {\it Theorem 5. } Given a NC manifold $M$ with metric $\g$,
there exists a unique map $\dds: \Xi_\st\times \Xi_\st\rightarrow \Xi_\st$ that satisfies the torsion free condition $T^\st(u,v)=0$ and the condition
\eq
\LL^\st_u\le v\otimes z , \g\res=\le \oR^\al(\dds_{\oR_\be(u)}\oR_\gamma(v))\ots \oR_\al\oR^\be\oR^\gamma (z),\g\res+
\le\oR^\al(v)\otimes_\st\dds_{\oR_\al(u)}(z), \g\res\label{mtricconsll} ~
\en
for all $u, v, z\in \Xi_\st$.
This map is smooth, in the sense that $\dds_uv$ is a smooth vector field if $u$ and $v$ are smooth. It is local in the sense that if $u=u'$ and $v=v'$ in an open $U\subset M$ then
$
\hat\dd^\st_u(v)=\hat\dd^\st_{u'}(v')
$ in $U\subset M$.

\begin{proof}
As in the undeformed case, use the
$\st$-symmetry of the metric (cf. \eqn{omompr}) and the torsion free condition $T^\st(u,v)=0$
 in order to rewrite
\eqn{mtricconsll}  as
\eqa
\LL^\star_u\le v\otimes z, \g\res&=&\le \oR^\be\oR^\gamma (z)\ots\dds_{\oR^\de\oR_\ga(v)}\oR_\delta\oR_\be(u) ,\g\res+\nn\\[.2em]
&& +\, \le  \oR^\be\oR^\ga (z)\ots ([\oR_\be(u),\oR_\ga(v)]_\st),\g\res+\le\oR^\ga(v)\otimes_\st \dds_{\oR_\ga(u)}z, \g\res\nn
\ena
Then sum and subtract the similar expression obtained by considering respectively the (undeformed) cyclic permutations $u,v,z\rightarrow v,z,u$ and
$u,v,z\rightarrow z,u,v$.
If we expand the $R$-matrix as $R=1\otimes 1+{\cal{O}}{(\la)}$, and the $\st$-Lie derivative as $\LL^\st_u=\LL_u+{\cal O (\la)}$,
where $\la$ is the deformation quantization parameter such that for $\la\rightarrow 0$
we have the undeformed product (for example for abelian twist $\la$ appears in 
 $\FF=e^{-{i\over 2}\la\theta^{ab}X_a\otimes X_b}$, and we have always included $\la$ in $\theta^{ab}$), the result is 
\eqa\label{qqqqq}
2\g(\dds_uv,z)&=&\LL_u\g(v,z)+\LL_v\g(z,u)-\LL_z\g(u,v)+
\g(v,[z,u])+
\g(z,[u,v])-
\g(u,[v,z])\nn\\
& &~~~~~~~~~~~~~~~~~~~~~~~~~~~~~~~~~~~~~~~~~~~~~+\la Fun(\dds, u,v,z,\g,\FF)
\ena
where $Fun(\dds, u,v,z,\g,\FF)$ is in particular a function of the connection $\dds$.
Now expand $\dds$ in power series of the noncommutativity parameter  $\la$,
$\dds=\dd+\la \dd_1+\la^2\dd_2\ldots$.
The $n^{\rm{th}}$-order term $\dd_n$ will depend, via \eqn{qqqqq}, from
the Lie derivative, $u,v,z, \g$ and $\dd_1,\ldots \dd_{n-1}$. Thus
\eqn{qqqqq} determines order by order in $\la$ the NC map $\dds_uv$ for any $u,v$.
Smoothness and locality properties are also immediate. 
\end{proof}

\section{Geometry and Gravity Solutions III:\\ Affine Killing Twists}
In this section,  as in the previous section,  we consider a general twist $\FF$  on a manifold $M$, and we study connections that are compatible with $\FF$. Their curvature and torsion tensors are undeformed. This way we  arrive at gravity solutions associated to a general twist, not necessarily of the abelian kind. 
\subsection{Killing and affine Killing vector fields}
Let us consider a (pseudo-)Riemannian manifold $M$ with metric $\g$. 
A Killing vector field $K$ is a vector field that leaves invariant the metric tensor,
\eq\label{Killing}
\LL_K\g=0~,
\en
i.e., for any $u,v\in \Xi$
\eq\label{LLKg}
\LL_K\,\g(u,v)=\g(\LL_K(u),v)+\g(u,\LL_K(v))~.
\en
Equivalently the (local) $1$-parameter group of diffeomorphisms associated with the
vector field $K$ consists of (local) isometries.

The Lie bracket of two Killing vector fields is again a Killing vector field
(indeed $\LL_{[K,K']}\g=\LL_K\LL_{K'}\g-\LL_{K'}\LL_{K}\g=0$). 
We thus have the Lie algebra of Killing vector fields. 

Let us now consider the unique torsion free metric compatible connection $\dd$ associated with $\g$ (the Levi-Civita or Riemannian connection).
It is easy to prove that the Lie derivative along a Killing vector field of the 
covariant derivative $\dd$ vanishes: 
\eq
\LL_K\dd=0~,\label{LLKDD}
\en
i.e., for any $u,v\in \Xi$\footnote{
{\sl Proof}. Recall that the Levi-Civita connection is uniquely determined by the condition, for all $u,v,z\in \Xi$,
\eq\label{ddLLg}
2\g(\dd_uv,z)=\LL_u\g(v,z)+\LL_v\g(z,u)-\LL_z\g(u,v)+
\g(v,[z,u])+
\g(z,[u,v])-
\g(u,[v,z])
\en
(that using local coordinates $x^i$ and the corresponding vector fields $\partial_i$ 
reads 
$2\Gamma_{ij}^k\g_{kl}=\partial_i\g_{jk}+\partial_j\g_{ki}-\partial_kg_{ij}$).
Acting with the Killing vector field $K$ on \eqn{ddLLg} and recalling \eqn{LLKg} 
and writing $\LL_K\LL_u\g(v,z)=\LL_{[K,u]}\g(v,z)+\LL_u\g(\LL_Kv,z)+\LL_u\g(v,\LL_Kz)$
we obtain \eqn{LLKdduv}.}

\eq\label{LLKdduv}
\LL_K(\dd_uv )=\dd_{[K, u]}v+\dd_u\LL_K v~,
\en
or equivalently, recalling that $\dd_uv=\le u, \dd v\re$, for any $v\in \Xi$ 
\eq\label{LLKddv}
\LL_K(\dd v )=\dd\LL_K v~.
\en

More generally we call a vector field $K$ an affine Killing vector field of $\dd$ if it is  compatible with the connection $\dd$ in the sense that \eqn{LLKDD} or \eqn{LLKdduv} holds (here $\dd$ is an arbitrary connection not necessarily the Levi-Civita one). Geometrically the flux associated to $K$ transforms 
parallel transported vector fields on a curve $\gamma$ into parallel transported vector fields on the push forward of $\gamma$.

If we consider the Levi-Civita connection then 
a special class of infinitesimal Killing vector fields is provided by homothetic 
Killing vector fields, i.e. vector fields $K$ that satisfy
\eq
\LL_K\g=c_{\,}\g
\en
for some constant $c$ (dependent on $K$).
Homothetic Killing vector fields form a Lie algebra.
It is easy to see that a homothetic Killing vector field is also an affine vector field
(hint: apply the Lie derivative $\LL_K$ to equation \eqn{ddLLg}).

We denote by $\trc$, $\rrc$ and $\ricc$ the 
 commutative torsion, curvature and Ricci curvature associated with $\dd$. 
If $\dd$ is the Levi-Civita connection of a (pseudo-)Riemannian manifold
then we further denote by 
 $\G$ and $\mathfrak{R}$  the Einstein tensor and the curvature scalar.

We recall that if $K$ is an affine Killing vector then  $\trc$, $\rrc$ and $\ricc$ are invariant under $K$,
\eqa
\LL_K\trc(u,v)&=&\trc(\LL_K u,v)+\trc(u,\LL_K v)~,~~\\
\LL_K\rrc(u,v,z)&=&\rrc(\LL_K u,v,z)+\rrc(u,\LL_K v,z)+\rrc(u,v,\LL_K z)~,~~\\
\LL_K\ricc(u,v)&=&\ricc(\LL_K u,v)+\ricc(u,\LL_K v)~,~~
\ena
(for a proof of this known result see e.g. \cite{KobNom}, Chapter 6).

We have introduced the three Lie algebras of  affine Killing vector fields, of  homothetic Killing vector fields and of Killing vector fields. Depending on the Riemannian manifold we are considering these three notions can coincide.  In particular we recall (see e.g. \cite{KobNom}, Chapter 6) that in an irreducible Riemannian manifold (i.e., a manifold whose holonomy group acts irreducibly) every infinitesimal affine transformation is homothetic. Moreover on a compact Riemannian manifold 
every affine Killing vector field is a Killing vector field. 

\subsection{Affine Killing twists}
We now consider a noncommutative deformation of a  manifold $M$ with connection $\dd$.  We  study the case
\eq
\FF\in U\hat g_K\otimes U\hat g_K~,\label{comptwist}
\en
where $\hat g_K$ is the Lie algebra of affine Killing vectors of the connection $\dd$. 
We thus relate the noncommutative structure of $M$ to the symmetries of the linear connection $\dd$ of $M$.

Later on we consider $\dd$ to be the Levi-Civita connection of the (pseudo-)Riemannian manifold $M$ with metric $\g$.

\noi {\it Theorem 6.~} There is a canonical $\st$-connection $\dd^\st$
associated to the $\st$-noncommutative manifold $M$ with connection $\dd$ and compatible twist $\FF$ as in \eqn{comptwist}.
The $\st$-connection $\dds$ is the undeformed connection $\dd$ itself,
\eq
\dd^\st=\dd~,
\en
where $\dds:\Xis\rightarrow \Omega_\st\otimes_\st \Xis$ while 
$\dd:\Xi\rightarrow \Omega\otimes \Xi$ and we use  that as vector spaces
$\Xis=\Xi$, $\Omega_\st=\Omega$ and $\Xis\otimes_\st\Omega_\st=\Xi\otimes\Om$.

The relation between the corresponding covariant derivatives is, for all $u\in \Xis$,
\eq
\dd_u^\st=\dd_{\of^\al(u)}\circ \of_\al~,\label{covderddsdd}
\en
where $\dds_u:\TT_\st\rightarrow \TT_\st$,
$~\dd_u:\TT\rightarrow \TT$ and we use  that as vector spaces $\TT_\st=\TT$ and $\Xis=\Xi$.
\sk
\begin{proof}
In order to show that $\dd$ is a noncommutative connection,
i.e., in order to show that \eqn{prop_Delta} holds,  we use  that the Lie derivative along an affine Killing vector commutes with the covariant derivative and the exterior derivative (cf. the derivation of  \eqn{dhgLeib}), and we recall that the action of the twist on tensors is via the Lie derivative. Therefore we have
\eqa
\dd(h\st v)=\dd(\of^\al(h)\of_\al(v))&=&
\d\of^\al(h)\otimes \of_\al(v)+\of^\al(h)\dd\of_\al(v)\nn\\
&=&\of^\al(\d h)\otimes \of_\al( v)+\of^\al(h)\of_\al (\dd v)\nn\\
&=&\d h\otimes_\st  v+h\st\dd v\nn~.
\ena

In order to prove relation \eqn{covderddsdd} we first observe that this relation holds when
$\dd_u^\st$ and $\dd_{\of^\al(u)}\circ \of_\al$ act on functions, indeed on functions $\dds_u=\LL^\st_u=\LL_{\of^\al(u)}\circ\of_{\al}=\dd_{\of^\al(u)}\circ \of_\al$. Similarly on vector fields,
\eq
\dd_u^\st v:=\le u,\dd^\st v\res=\le \of^\al(u),\of_\al(\dds v)\re= 
\le \of^\al(u),\dds \of_\al(v)\re=\dd_{\of^\al(u)} \of_\al(v)~. 
\en
Next we show that $\dd_{\of^\al(u)}\circ \of_\al$ satisfies the same deformed Leibniz rule as $\dd_u^\st$. We notice that in $Ug\otimes Ug\otimes Ug$ we have
\eqa
\of^\al_2\of_\be\otimes\of^\al_1\of^\be\otimes \of_\al
&=&
\of^\al_1\of_\be\otimes\of^\al_2\of^\be\otimes \of_\al\nn\\[.2em]
&=&
\of^{\al}_1 \of^\epsi \f^\sigma\of_\be\otimes \of^\al_2\of_\epsi\f_\sigma\of^\be \otimes \of_\al\nn\\[.2em]
&=&
\of^\al\f^\sigma\of_\be \otimes\of_{\al_1}\of^\epsi\f_\sigma\of^\be \otimes\of_{\al_2}\of_\epsi\nn\\[.2em]
&=&
\of^\al\oR^\ga\otimes\of_{\al_1}\of^\epsi\oR_\ga\otimes \of_{\al_2}\of_\epsi 
\label{technical}
\ena
where in the first line we have used
that the undeformed coproduct is cocommutative (and therefore
$\of^\al_2\otimes\of^\al_1\otimes \of_\al=\of^\al_1\otimes\of^\al_2\otimes\of_\al$   ), in the second line we inserted the identity in the form $1\otimes 1=\FF\FF^{-1}=\of^\epsi\f^\sigma\otimes\of_\epsi\f_\sigma$, in the third line we used \eqn{ass}, and in the fourth line we recalled that $\RR^{-1}=\FF\FF_{21}^{-1}$.
We then  compute
\eqa
\dd_{\of^\al\!(u)}\of_\al(v\otimes_\st z)&=&
\dd_{\of^\al\!(u)}\big(\of_{\al_1}\of^\be(v)\otimes\of_{\al_2}\of_\be(z)\big)\nn\\[.2em]
&=&
\dd_{\of^\al_1\of^\be(u)}\big(\of^\al_2\of_\be(v)\otimes\of_\al(z)\big)\nn\\[.2em]
&=&\dd_{\of^\al_1\of^\be\!(u)}\big(\of^\al_2\of_\be(v)\big)\otimes\of_\al(z)
+\of^\al_2\of_\be(v)\otimes\dd_{\of^\al_1\of^\be(u)}\of_\al(z)\nn\\[.2em]
&=&\le\of^\al_1 \of^\be(u),\of^\al_2\of_\be(\dd v)\re\otimes\of_\al(z)
+\of^\al\oR^\ga(v) \otimes\le\of_{\al_1}\of^\epsi\oR_\ga(u), \dd_{}\of_{\al_2}\of_\epsi (z)\re
\nn\\[.2em]
&=&\of^\al\le u,\dd v\res\otimes\of_\al( z) 
+\of^\al\oR^\ga(v) \otimes\of_\al\le\oR_\ga(u), \dd z\res\nn\\[.2em]
&=&
\le u,\dd v\res\otimes_\st z
+\oR^\ga(v)\otimes_\st\le\oR_\ga(u),\dd z\res\nn\\[.2em]
&=&
\dds_u v\otimes_\st z 
+\oR^\ga(v)\otimes_\st \dd^\st_{\oR_\ga(u)} z
\label{leibddu}
\ena
where in the second line we have used \eqn{ass}, and in the fourth line we recalled \eqn{technical}.
This expression coincides with \eqn{Leibdds} for affine Killing twists. This can be seen by repeatedly applying \eqn{ass} and cocommutativity of the undeformed coproduct, or also by considering the following equalities,
\eqa
\oR^\al(\dds_{\oR_\be(u)}\oR_\gamma(v))\ots \oR_\al\oR^\be\oR^\gamma (z)&=&
\oR^\al\le{\oR_\be(u)}, \dds\oR_\gamma(v)\res\ots \oR_\al\oR^\be\oR^\gamma (z)\nn\\
&=&
\le\oR^\al_1\oR_\be(u), \oR^\al_2\dds\oR_\gamma(v)\res\ots \oR_\al\oR^\be\oR^\gamma (z)\nn\\
&=&
\le\oR^\al_1\oR_\be(u), \dds\oR^\al_2\oR_\gamma(v)\res\ots \oR_\al\oR^\be\oR^\gamma (z)\nn\\
&=&
\le\oR^\al\oR_\be(u), \dds\oR^\sigma\oR_\gamma(v)\res\ots \oR_\sigma\oR_\al\oR^\be\oR^\gamma (z)\nn\\
&=&
\le u, \dds v\res\ots z\nn\\
&=&
\dds_u v\ots z\,,
\ena
where in the fourth line we used the $R$-matrix property 
$(\Delta\ots id){\cal R}^{-1}={\cal R}^{-1}_{23}{\cal R}^{-1}_{13}$, i.e. $\oR^\al_1\ots\oR^\al_2\ots\oR_\al=\oR^\al\ots \oR^\sigma\ots \oR_\sigma\oR_\al$, that follows from \eqn{ass}, and in the fifth line we used the triangularity property ${\cal R}_{12}={\cal R}^{-1}_{21}$, that immediately follows from the definition of ${\cal R}$.

Since the Lie derivative and the covariant derivative commute with the contraction operator $\le~,~\re$, we also have $\dd_{\of^\al(u)} \of_\al\,\om=\dds_u\om$ for any 1-form (the proof is similar to \eqn{leibddu} just consider $\le v,\om\res$ rather than $v\otimes_\st z$; then recall \eqn{compleredd}). 
This implies that the deformed Leibniz rule property \eqn{leibddu} holds also if $v$ and/or $z$ are 1-forms. Iterated use of this property then shows that $\dd_{\of^\al(u)}\circ \of_\al=\dds_u$ on any tensor field and for any vector field $u\in \Xis$. 
\end{proof}

\sk
\noi {\it Theorem 7.~} The $\st$-torsion, $\st$-curvature and $\st$-Ricci tensors
of the connection $\dds=\dd$ are the undeformed ones,
\eq
\tr=\trc~,~~ \rr=\rrc~,~~\ric=\ricc~.
\en
These equalities holds when we consider $\tr,\rr, \ric$ as tensors, and we use that  the noncommutative and commutative tensor spaces are equal as vector spaces,
$\TT_\st=\TT$.  When we consider $\tr,\rr, \ric$ as multilinear operators we have,  due to invariance of $\trc,\rrc, \ricc$ under affine Killing vector fields,
\eq
\le u\otimes_\st v , \tr \res = \le u\otimes_\st v,\trc\re\,,~\le u\otimes_\st v\otimes_\st z ,\rr\res=\le u\otimes_\st v\otimes_\st z ,\rrc\re\,,~
\le u\otimes_\st v ,\ric\res=\le u\otimes_\st v ,\ricc\re\nn
\en
for all $u,v,z\in\Xis$.
\begin{proof}
In order to prove that $\tr=\trc$ we show that $\le u\otimes_\st v , \tr \res = \le u\otimes_\st v,\trc\res$ 
for all $u,v\in \Xis$,
\eqa
\le u\otimes_\st v, \tr\res=\tr(u,v)&=&\le u,\dd v\res-\le{\oR^{\alpha}(v)},\dd\oR_{\alpha}(u)\res
-[u,v]_{\star}\nn\\[.2em]
&=&\le\of^\al (u),\of_\al(\dd v)\re-\le{\of^\be\oR^{\alpha}(v)},\of_\be(\dd\oR_{\alpha}(u))\re
-[\of^\al (u),\of_\al (v)]\nn\\[.2em]
&=&\le\of^\al (u),\dd\of_\al( v)\re-\le \of_\al(v),\dd\of^{\alpha}(u)\re
-[\of^\al (u),\of_\al (v)]\nn\\[.2em]
&=&\trc(\of^\al(u),\of_\al(v))\nn\\
&=&\le u\otimes_\st v,\trc\re\nn\\
&=&\le u\otimes_\st v,\trc\res\nn
\ena
where we used the definition \eqn{defUR} of the $R$-matrix and, in the last equality,
that the torsion tensor $\trc$ is invariant under affine Killing vector fields.

We similarly prove $\rr=\rrc$, 
\eqa
\le u\otimes_\st v\otimes_\st z,\rr\res&=& 
\dds_u\dds_v z-\dds_{\oR^\ga(v)}\dds_{\oR_\ga(u)}z-\dds_{[u,v]_\st}z\nn\\[.2em]
&=&\le u,\dd\le v,\dd z\res\res-\le\oR^\ga(v),\dd\le\oR_\ga(u),\dd z\res\res-\le[u,v]_\st,\dd z\res\nn\\[.2em]
&=&\le \of^\al_1\of^\be(u),\dd\le \of^\al_2\of_\be(v),\dd \of_\al(z)\re\re
-\le \of^\al_2\of_\be(v),\dd\le\of^\al_1\of^\be(u),\dd\of_\al (z)\re\re \nn\\
&&~~~~~~~~~~~~~~~~~~~~~~~~~~~~~~~~~~~~~~~~~~~~-\le[\of^\al_1\of^\be(u),\of^\al_2\of_\be(v)],\dd \of_\al(z)\re\nn\\
&=&\le u\otimes_\st v\otimes_\st z, \rrc\re\nn\\[.2em]
&=&\le u\otimes_\st v\otimes_\st z, \rrc\res
\ena
where in the third line we used \eqn{ass} and \eqn{technical}.

In order to show that $\ric=\ricc$ we write the identity operator on the space of 1-forms in two equivalent ways,
\eq
{\rm id}=\check\theta^i\le\check e_i,(\;.\;)\re=\theta^i\st\le e_i, (\;.\;)\res
\label{idbbs}
\en
where $\check e_i$ and $\check\theta^i$ are dual bases while $e_i$ and $\theta^i$
are $\st$-dual bases, 
$
\,\le\check e_i,\check \theta^j\re=\delta^j_i~,~~
\le e_i,\theta^j\res=\delta^j_i~.
$ 
The equalities \eqn{idbbs} follow immediately by decomposing  a 1-form as
$\om=\check\theta^j\om_j=\theta^j\st\om^\st_j$.
These equalities imply that on any tensor $\tau\in \TT^{1,1}$ 
\eq
\le \theta^i,\le e_i , \tau\res\res=\le\check\theta^i , \le\check e_i,\tau\re\re~.
\label{idbbs2}
\en
Indeed (locally)  write $\tau=\tau^j\otimes_\st e_j$, where $\tau^j$ are 1-forms; it then follows
$$
\le \theta^i,\le e_i ,\tau^j\otimes_\st e_j\res\res=\le\theta^i , \le e_i,\tau^j\res\st e_j\res=
\le \theta^i\st\le e_i , \tau^j\res , e_j\res = \le \tau^j , \e_j\res
=\le\of^\al(\tau^j) , \of_\al(e_j)\re$$
$$
\le \check\theta^i,\le\check e_i ,\tau^j\otimes_\st e_j\re\re=
\le \check\theta^i,\le\check e_i ,\of^\al(\tau^j) \otimes \of_\al( e_j)\re\re=
\le\check\theta^i , \le \check e_i,\of^\al(\tau^j)\re ,\of_\al(e_j)\re=
\le\of^\al(\tau^j) , \of_\al(e_j)\re
$$
We then compute, for all $u,v\in\Xis$,
\eqa
\le u\otimes_\st v,\ric\res&=&\le \theta^i,\le e_i\otimes_\st u\otimes_\st v,\rrc\res\res\nn\\[.2em]
&=&\le \theta^i,\le e_i , \le u\otimes_\st v,\rrc\res\res\res\nn\\[.2em]
&=&\le\check\theta^i , \le\check e_i,\le u\otimes_\st v,\rrc\res\re\re\nn\\[.2em]
&=&\le\check\theta^i , \le\check e_i,\le u\otimes_\st v,\rrc\re\re\re\nn\\[.2em]
&=&\le\check\theta^i , \le\check e_i\otimes  \of^\al(u)\otimes \of_\al(v),\rrc\re\re\nn\\[.2em]
&=&\le \of^\al(u)\otimes \of_\al(v),\ricc\re\nn\\[.2em]
&=&\le u\otimes_\st v,\ricc\res
\ena
where in the second line we used the $\st$-pairing property \eqn{nestedpairing}, in the third line  property \eqn{idbbs2} with $\tau=\le u\otimes_\st v,\rrc\res$, and in the fourth and last lines the invariance of
$\rrc$ and $\ricc$ under affine Killing vector fields.
\end{proof}

\subsection{Gravity solutions III}

We now consider a (pseudo-)Riemaniann manifold $M$ with metric $\g$ and associated Levi-Civita connection $\dd$. If $\g$ is positive definite 
(and if any two points of $M$ can be connected by a geodesic) then
we have the de Rham decomposition (see for ex. ref. \cite{DeRHam}), $M=M_1\times\ldots M_p$, where $M_1$ is the Euclidean space ${\mathbb{R}}^m$, $m\geq 0$, and $M_i$, $i=2,\ldots p$ are irreducible Riemannian manifolds not isometric to the real line. 
The metric $\g$ is the direct sum of the standard Euclidean metric of ${\mathbb{R}}^m$ and of the metrics $\g_i$ of $M_i$.  
In this setting an affine Killing vector acts on each manifold $M_i$ as an
 homothetic Killing vector ($\LL_K\g_i=c_i \g_i$), see Theorem 3.6 in \cite{KobNom}. 

If $M$ has Lorentzian (or more in general indefinite) signature, then, given a decomposition $M= M_1\times\ldots M_p$, we consider affine Killing vector fields $K$ that when restricted to 
each $M_i$,  $i=1,2,\ldots p$ act as homothetic Killing vector fields. These vector fields form a Lie subalgebra of the Lie algebra $\hat g_K$ of affine Killing vector fields of $M$. 
We denote it by $g_{_{hK}}$. 

In this section we consider metrics $\g$ that are compatible with the twist $\FF$ in the sense that 
\eq
\FF\in Ug_{_{hK}}\otimes Ug_{_{hK}}~.\label{comptwists}
\en
If this is the case we have
\sk
\noi {\it Theorem 8.~} The $\st$-Levi-Civita connection $\dd^\st$
associated with the $\st$-noncommutative manifold $M$ with twist $\FF$
and compatible metric $\g$   as in \eqn{comptwists}
is the usual Levi-Civita connection of the commutative manifold $M$ with metric $\g$,
\eq
\dd^\st=\dd~,
\en
where $\dds:\Xis\rightarrow \Omega_\st\otimes_\st \Xis$ while 
$\dd:\Xi\rightarrow \Omega\otimes \Xi$ and we use  that as vector spaces
$\Xis=\Xi$, $\Omega_\st=\Omega$ and $\Xis\otimes_\st\Omega_\st=\Xi\otimes\Om$.

\begin{proof}
Because of Theorem 6 we just have to prove the compatibility of $\dds$ with the metric tensor.
Let $\g=\g_1+\g_2+ \ldots \g_p$ be the metric on $M=M_1\times M_2\times\ldots M_p$, and $v=v^1+ v^2+\ldots v^p$ a vector field. Then the Levi-Civita connection is the direct sum of the Levi-Civita connections on $M_1, M_2,\ldots M_p$, 
$\dd_v=\dd^1_{v^1}+\dd^2_{v^2} +\ldots \dd^p_{v^p}$. From the very definition of the Lie algebra $g_{_{hK}}$ and from \eqn{comptwists} it follows
that for each index $\al$, $\of_\al \g=c^1_\al\g_1+ c^2_\al\g_2 +\ldots c^p_\al\g_p$
with $c^1_\al, c^2_\al,\ldots c^p_\al$ constant coefficients. Finally we have,
\eq
\dds_u\g=\dd_{\of^\al(u)}\of_\al \,\g=\sum_{i=1}^p\dd^i_{{\of^\al(u)}^i}\,c^i_\al\g_i=0~.
\en
\end{proof}

We now apply Theorem 7 and conclude that the torsion, curvature and Ricci tensors 
of the connection $\dds=\dd$ are the undeformed ones. Therefore,
\sk
\noi {\it Corollary 2.~} If  $\g$ is a commutative Einstein metric for the manifold $M$ and \eqn{comptwists} holds, then $\g$ is also a noncommutative Einstein metric.
$~~~~~~~~~~~~~~~~~~~~~~~~~~~~~~~~~~~~~~~~~~~~~~~~~~~~~~~~~~~~~~~~~~~~~~~~${\small{$\square$}}

\sk
\sk
For example let's consider the Connes-Landi 4-sphere \cite{Connes-Landi}.
It is obtained from an abelian Drinfeld twist constructed with Killing vector fields.
It therefore satisfies condition \eqn{comptwists}. It is a noncommutative Einstein space
with noncommutative connection, curvature  and Ricci tensors equal to the undeformed ones.

Explicitly the usual 4-sphere is the subspace of $\mathbb{R}^5$ defined by
$\sum_{i=1}^5{X^i}^2=1$, or, using the complex coordinates $x^1=(X^1-iX^2)/\sqrt{2}\,,~x^2=(X^4-iX^5)/\sqrt{2}\,,~x^3=X^3\,,~x^4=\overline{x^2}\,,~x^5=\overline{x^1}$, by 
$
2x^1x^5+2x^2x^4+x^3x^3=1$. The twist is 
\eq\FF=e^{{-i\over 2}\la [(x^1\partial_1-x^5\partial_5)\otimes (x^2\partial_2-x^4\partial_4)
- 
(x^2\partial_2-x^4\partial_4)\otimes (x^1\partial_1-x^5\partial_5)]}~.
\en
The $x^i$ coordinates satisfy the noncommutative 4-sphere relations (cf. for example \cite{Aschieri-Bonechi}):
\eqa
2x^1\st x^5+2x^2\st x^4+x^3\st x^3=1~~,~~~x^3\st x^i=x^i\st x^3~~ (i=1,2,4,5)\nn\\[.2em]
x^1\st x^2=q x^2\st x^1~,~~x^1\st x^4=q^{-1} x^4\st x^1~,~~x^1\st x^5= x^5\st x^1~,\nn\\[.2em]
x^2\st x^5=q x^5\st x^2~,~~x^4\st x^5=q^{-1} x^5\st x^4~,~~x^2\st x^4= x^4\st x^2~,\nn
\ena
where $q=e^{i\la}$.

\end{document}